\documentclass{aastex}
\usepackage[utf8]{inputenc}
\usepackage{graphicx}	
\usepackage{natbib}
\usepackage{appendix}

\title{SDSS-IV MaNGA: Environmental dependence of the Mgb/$\langle$Fe$\rangle$-$\sigma_*$ relation for nearby galaxies}
\author{Zheng Zheng\altaffilmark{1,2}, 
Cheng Li\altaffilmark{3}, 
Shude Mao\altaffilmark{3,1}, 
Huiyuan Wang\altaffilmark{4}, 
Chao Liu\altaffilmark{1}, 
Houjun Mo\altaffilmark{3,5},  
Zhen Yuan\altaffilmark{6},  
Claudia Maraston\altaffilmark{7}, 
Daniel Thomas\altaffilmark{7}, 
Renbin Yan\altaffilmark{8}, 
Kevin Bundy\altaffilmark{9}, 
R. J. Long\altaffilmark{1,10},
Taniya Parikh\altaffilmark{7}, 
Grecco Oyarz\'un\altaffilmark{9}, 
Dmitry Bizyaev\altaffilmark{11},
Ivan Lacerna\altaffilmark{12,13}
}
\altaffiltext{1}{National Astronomical Observatories, Chinese Academy of Sciences, A20 Datun Road, Chaoyang District, Beijing 100101, China}
\altaffiltext{2}{CAS Key Laboratory of FAST, NAOC, Chinese Academy of Sciences}
\altaffiltext{3}{Tsinghua Center for Astrophysics and Department of Physics, Tsinghua University, Beijing 100084, China}
\altaffiltext{4}{Key Laboratory for Research in Galaxies and Cosmology, Department of Astronomy, University of Science and Technology of China, Hefei, Anhui 230026, China}
\altaffiltext{5}{Department of Astronomy, University of  Massachusetts Amherst, MA 01003, USA}
\altaffiltext{6}{Shanghai Astronomical Observatory, Chinese Academy of Sciences, Shanghai 200030, China} 
\altaffiltext{7}{Institute of Cosmology and Gravitation, University of Portsmouth, Dennis Sciama Building, Burnaby Road, Portsmouth, PO1 3FX, UK}
\altaffiltext{8}{University of Kentucky, Lexington, KY 40506, USA} 
\altaffiltext{9}{Department of Astronomy and Astrophysics, University of California, 1156 High Street, Santa Cruz, CA 95064}
\altaffiltext{10}{Jodrell Bank Centre for Astrophysics, School of Physics and Astronomy, The University of Manchester, Oxford Road, Manchester M13 9PL, UK}
\altaffiltext{11}{Apache Point Observatory and New Mexico State University, P.O. Box 59, Sunspot, NM, 88349-0059, USA}
\altaffiltext{12}{Instituto de Astronom\'ia, Universidad Cat\'olica del Norte, Av. Angamos 0610, Antofagasta, Chile}
\altaffiltext{13}{Instituto Milenio de Astrof\'isica, Av. Vicu\~na Mackenna 4860, Macul, Santiago, Chile}

\begin{document}

\begin{abstract}
We use a sample of $\sim3000$ galaxies  from the MaNGA MPL-7 internal data release to study the $\alpha$ abundance distribution within low-redshift galaxies. We use the Lick index ratio Mgb/$\langle$Fe$\rangle$ as an $\alpha$ abundance indicator to study relationships between the $\alpha$ abundance distribution and galaxy properties such as effective stellar velocity dispersion within $0.3$ effective radii ($\sigma_*$), galaxy environment, and dark matter halo formation time ($z_f$). We find that
(1) all galaxies show a tight correlation between Mgb/$\langle$Fe$\rangle$  and $\sigma_*$; 
(2) `old' (H$\beta<3$) low-$\sigma_*$ galaxies in high local density environment and inner regions within galaxy groups are enhanced in Mgb/$\langle$Fe$\rangle$, while `young' (H$\beta>3$)  galaxies and high-mass galaxies show no or less  environmental dependence; 
(3) `old' galaxies with high-$z_f$ show enhanced Mgb/$\langle$Fe$\rangle$ over low- and medium-$z_f$;
(4) Mgb/$\langle$Fe$\rangle$ gradients are close to zero and show  dependence on $\sigma_*$ but no obvious dependence on the environment or $z_f$. 
Our study indicates that stellar velocity dispersion or galaxy mass is the main parameter driving the Mgb/$\langle$Fe$\rangle$ enhancement, although environments appear to have modest effects, particularly for low- and medium-mass galaxies.
\end{abstract}


\maketitle

\section{Introduction}
The $\alpha$-to-iron ratio ([$\alpha$/Fe]) is an important indicator for star formation histories because $\alpha$-elements ($\rm O, Ne, Mg, Si, Ca, Ti$) are mostly produced in core-collapse supernovae, whose progenitors are high-mass stars, while irons (Fe) are mostly produced by type Ia supernovae, whose progenitors are low-mass compact stars. Therefore, a galaxy formed in a single burst with a top-heavy initial mass function (IMF) will be enhanced in [$\alpha$/Fe] in comparison with a galaxy formed with an extended star formation history and a bottom-heavy IMF \citep[][and references therein]{Thomas et al.1998, Thomas1999, Matteucci1994}.

There have been many studies on the [$\alpha$/Fe] ratios of nearby galaxies and a major finding is that the [$\alpha$/Fe] ratios of early type galaxies are best correlated with stellar velocity dispersion $\sigma$ \footnote{Here we use $\sigma$ to denote velocity dispersion measured from the literature, which are mostly measured using single fiber or long slit spectra and has various aperture sizes and definitions. We use $\sigma_*$ (in later text) to denote effective velocity dispersion measured within 0.3 effective radius using IFU data (see detailed definition of $\sigma_*$ is in Section \ref{def_vd}).} \citep[e.g.][]{Worthey et al.1992, Trager et al.2000, Thomas et al.2005, Sanchez-Blazquez et al.2006, Sansom et al.2008, Thomas et al.2010, Kuntschner et al.2010, Johansson et al.2012, Liu et al.2016}. Studies which do not have morphological cut \citep[e.g.][]{Proctor et al.2002, Smith et al.2007, Scott et al.2017} find similar trends for late type galaxies, albeit using smaller sample sizes. 

The [$\alpha$/Fe]-$\sigma$ correlation implies that high-$\sigma$ (mostly massive) galaxies have a shorter star formation time scale and quench more sharply than low-$\sigma$ (mostly low-mass) galaxies. If the star formation is quenched in less than $\sim$ 1Gyr, later type Ia supernovae explosions will not be able to cause iron enrichment for stellar populations. 
Quenching could be due to internal processes such as supernovae \citep{DekelSilk1986} or AGN feedback \citep{King2003, Segers et al.2016}. 
Detailed explanation for  [$\alpha$/Fe]-$\sigma$ relation includes IMF variation, stellar yields, inflow/outflow rates and other galaxy evolution parameters \citep[e.g.][]{Pipino et al.2009, Arrigoni et al.2010, Yates et al.2013}. However, the effects of many parameters are degenerate. Also, high-mass galaxies are more likely to have a bottom-heavy IMF \citep[Zhou et al. submitted;][]{Cappellari et al.2012, Li et al.2017, Parikh et al.2018}, which results in an opposite  [$\alpha$/Fe]-$\sigma$ relation comparing to the observations because a bottom-heavy IMF leads to relatively fewer massive stars and thus produces less $\alpha$-elements as a result of fewer core collapse supernovae.

Environmental effects, such as harassment \citep[e.g.][]{moore98}, strangulation \citep[e.g.][]{balogh00, Peng et al.2015}, or gas stripping \citep[e.g.][]{gg72}, 
could also quench star formation and thus enrich [$\alpha$/Fe], especially for low-mass galaxies. Gas accretion from galaxy surroundings \citep{Larson et al.1980} can, in the other hand, cause continuous star formation and decrease the $\alpha$-to-iron ratio. Many observational studies using galaxies in clusters \citep[e.g.][]{Smith et al.2007, Michielsen et al.2008, Chilingarian et al.2008, Matkovic et al.2009, Spolaor et al.2010, Liu et al.2016a} and poor groups \citep[e.g.][]{Annibali et al.2011} show that galaxies in dense regions are indeed enhanced in [$\alpha$/Fe]. While some other studies \citep[e.g.][]{Sanchez-Blazquez et al.2003, Annibali et al.2007, Paudel et al.2011, McDermid et al.2015, Thomas et al.2010} claim little environmental effects on [$\alpha$/Fe] enhancement for early type galaxies. 

Recently,  \citet{Liu et al.2016} compiled measurements from the literature of [$\alpha$/Fe] and the Lick index ratios Mgb/$\langle$Fe$\rangle$ of thousands of early type galaxies in different environments and found that the low-$\sigma$ end of the [$\alpha$/Fe]-$\sigma$ relation has larger intrinsic scatters and is dependent on environment: low-$\sigma$ early type galaxies in high density clusters are enhanced in Mgb/$\langle$Fe$\rangle$ and [$\alpha$/Fe]. However, these measurements are from different telescopes and instruments and  have different systematic offsets and different apertures. Also, most previous works \cite[except for][]{Kuntschner et al.2010, Greene et al.2015, Scott et al.2017} focus on the central part of early type galaxies. Galaxy outskirts are more likely to be affected by interactions with nearby galaxies and thus experience different star formation histories \citep[e.g.][]{Bakos et al.2008, Ruiz-Lara et al.2016}. 
Furthermore, since galaxies are formed in dark matter halos and could experience many mergers during their formation, the distribution of [$\alpha$/Fe] could also be affected by halo merger histories. Therefore, it is important to study the dependence of [$\alpha$/Fe]-$\sigma$ relation on environments and formation histories using a uniform survey with spectroscopic coverage to galaxy outskirts. 

Thanks to the Sloan Digital Sky Survey (SDSS) Mapping Nearby Galaxies at APO \citep[MaNGA,][]{Bundy et al.2015} project, we are able to obtain integrated field unit (IFU) spectra and thus maps of $\alpha$- and iron-abundance sensitive spectral indices  for a large sample of galaxies. We therefore can study  $\alpha$-to-iron ratio distributions within different types of galaxies and explore their environmental dependence. In this paper, we use environment information derived through a method based on reconstructed density field \citep{Wang et al.2009, Wang et al.2012, Wang et al.2016} instead of neighboring galaxies. \citet{Wang et al.2016} also provides us with the halo formation redshift for individual galaxies, which enables us to investigate the effect of formation histories. We also explore the effects of central/satellite galaxies and projected halo-centric radius (projected distance between the satellite galaxy to the group center) using information from \citet{Yang et al.2007}. 

The outline of the paper is as follows: we describe the data and briefly introduce the measurements of spectral indices and different environment indicators  in Section \ref{datasample};
 we then present our results in Section \ref{results} and discuss them in Section \ref{discussion}; we summarize our conclusions in Section \ref{summary}. A discussion on the environmental dependence of [$\alpha$/Fe] is in  Appendix \ref{appendix}.

\section{Data and Methodology}
\label{datasample}

\subsection{MaNGA sample}

MaNGA \citep{Bundy et al.2015} is an IFU survey targeting about 10 000 nearby galaxies selected from the SDSS-IV \citep{Blanton et al.2017, Gunn et al.2006}. The wavelength coverage is between 3600 and 10300 \AA  \ with a spectral resolution $R \sim 2000$ \citep{Drory et al.2015, Smee et al.2013}. 
The MaNGA sample is selected so that the number density of galaxies is roughly independent of the $i$-band absolute magnitude $M_i$. 
Each galaxy is covered by the IFU out to either 1.5 effective radii ($R_e$) or 2.5$R_e$ \citep{Yan et al.2016b, Wake et al.2017} and the sizes of the IFUs vary for different galaxies from $12\arcsec$ for a 19-fiber IFU to $32''$ for a 127-fiber IFU \citep{Drory et al.2015, Law et al.2015}.
The observed data are reduced by the MaNGA data reduction pipeline \citep[DRP;][]{Law et al.2016, Yan et al.2016a} and the final products of the DRP are datacubes, which are composed of `spaxels' with an angular size of  $0.5''$x$0.5''$, as well as row stacked spectra.

\begin{figure}
\begin{center}
    \includegraphics[scale=0.5]{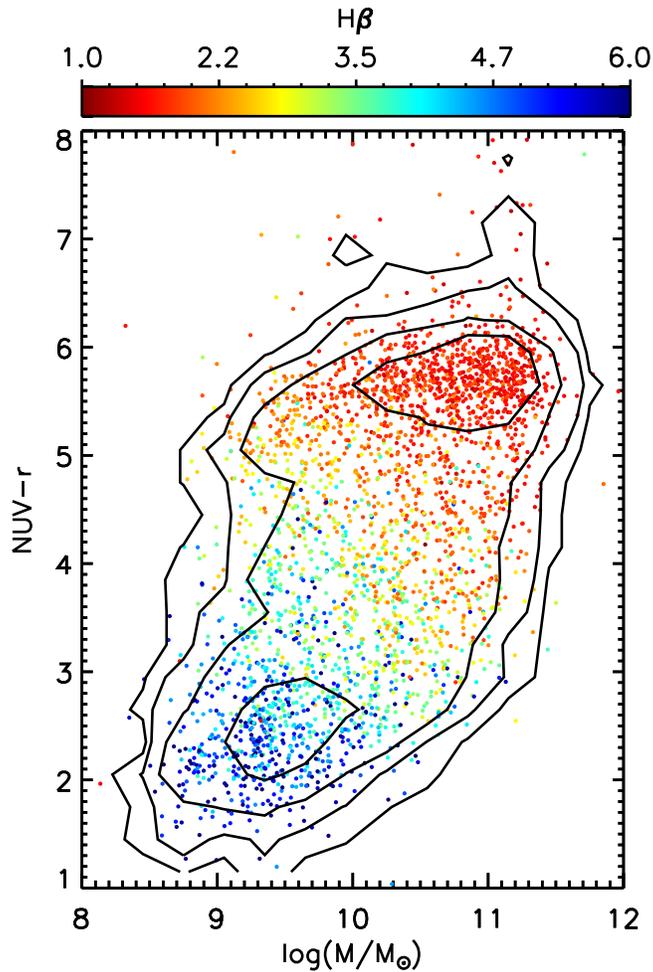}
    \caption{NUV-r color versus stellar mass for our sample, which includes 2997 galaxies. Each dot is a galaxy, color-coded by the average H$\beta$ index value (cf. Section \ref{lickindx} for details) within 0.3$R_e$ of each galaxy. The black contours show the mass-color distribution of the whole MaNGA MPL-7 sample (4621 galaxies).}
    \label{fig:mass_nuvr}
  \end{center}
\end{figure}

In this work, we use datacubes from the MaNGA internal data release MPL-7, which is equivalent to the SDSS DR15 public release \citep{Aguado et al.2019}. The MPL-7 sample contains 4621 galaxies and has a redshift range $0.01<z<0.15$ with a peak around $z=0.03$. 
We exclude galaxies with obvious interaction features (125 galaxies),
galaxies with  unreliable velocity dispersion measurements due to very low signal-to-noise ratios (468 galaxies)
and galaxies with bad background subtractions (e.g. obvious spikes) around 5000\AA or very broad emission lines (identified by eye, 1031 galaxies). The total number of galaxies used in this work is 2997. We do not have morphological cut and we include not only red sequence galaxies but also blue sequence and green valley galaxies (see stellar mass versus NUV-r color diagram in Fig. \ref{fig:mass_nuvr}). 

\subsection{Spectral index measurements}
\label{lickindx}
Absorption line indices have been widely used for studies of stellar population parameters, especially  for $\alpha$ abundance \citep[e.g.][]{Worthey1994, Thomas et al.2005, Thomas et al.2011}. There are many $\alpha$-element sensitive and iron-sensitive indices within the MaNGA wavelength coverage. Within these indices, H$\beta$, Mgb, Fe5270 and Fe5335 (see below for definitions) have been proven to be the strongest and best set of spectral indices  for studying $\alpha$ abundances \citep[e.g.][]{Thomas et al.2011, Liu et al.2016}. H$\beta$ is sensitive to stellar age and has a negative correlation with age; Mgb is most sensitive to magnesium, which is an important $\alpha$-element; Fe5270 and Fe5335 are most sensitive to iron abundance \citep{Thomas et al.2003, Thomas et al.2011}. We therefore focus on these four spectral indices in this paper. 

We follow the definitions of spectral indices from \citet{Trager et al.1998}. Detailed bandpass and pseudo-continua wavelength ranges for the four spectral indices are reproduced in Table \ref{sindx_def}. Uncertainties in the spectral index values are estimated following \citet{Cardiel et al.1998}.

\begin{table*}
\caption{Spectral index definitions used in this paper are from \citet{Trager et al.1998}. The bandpass is the wavelength range covering the absorption lines and the blue and red continuum are wavelength ranges around the bandpass for fitting the continuum slope within the bandpass. The wavelengths are measured in air and all numbers are in unit of \AA. }
\begin{center}
\begin{tabular}{cccc}
\hline
\hline
Name & Index Bandpass & Blue continuum & Red continuum \\
\hline
H$\beta$        & 4847.875 - 4876.625 &   4827.875 - 4847.875   & 4876.625 - 4891.625  \\
Mgb        & 5160.125 - 5192.625 &   5142.625 - 5161.375   & 5191.375 - 5206.375 \\
Fe5270     & 5245.650 - 5285.650 &   5233.150 - 5248.150   & 5285.650 - 5318.150  \\  
Fe5335     & 5312.125 - 5352.125 &   5304.625 - 5315.875   & 5353.375 - 5363.375   \\
\hline
\hline
\label{sindx_def}
\end{tabular}
\end{center}
\end{table*}%

We measure these four spectral indices for all spaxels with signal-to-noise ratios (SNR) larger than 10. 
For studying stellar population related science using spectral indices, the threshold requirement for spectral SNR is very high ($\rm SNR \gtrsim 50-100$).
Therefore, we divide each galaxy into three radial bins to enhance SNR: central ($r<0.3R_e$), intermediate ($0.3R_e<r<0.7R_e$), and outer ($0.7R_e<r<1.3R_e$) regions, where $r$ is semi-major axis of an elliptical aperture. We then calculate average values of the spectral indices within each radial bin. The  averaged spectral index values are similar to values derived using each radial bin's stacked spectra, which usually have $\rm SNR \gtrsim 50-100$. 

\begin{figure} 
\begin{center}
\includegraphics[scale=0.6]{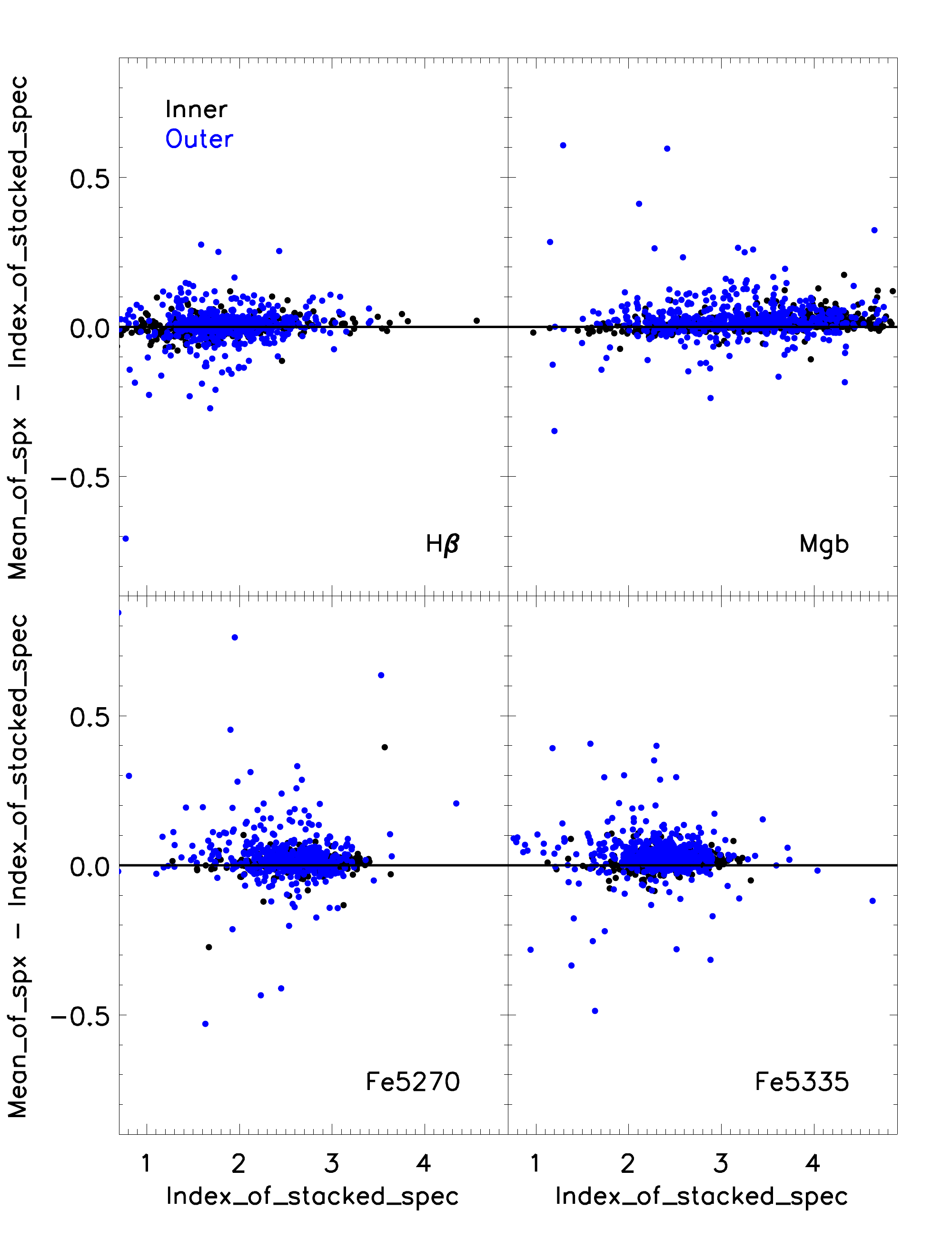}  
\caption{ Comparison of spectral indices derived using stacked spectra (Index\_of\_stacked\_spec) and mean values of spectral indices of individual spaxels (mean\_of\_spx).  
This plot is made using 500 galaxies  randomly selected from our sample.  The four panels show four different spectral indices respectively. 
The black dots are for inner regions of the galaxies and blue dots are for outer regions of galaxies.  The outliers in the upper-right (Mgb) and lower-right (Fe5335) panels are caused by bad stacked spectra. }
\label{compare_mean_vs_stack}
\end{center}
\end{figure}

A comparison of spectral indices derived using stacked spectra versus average spectral index values of individual spaxels is shown in Fig. \ref{compare_mean_vs_stack}. As can be seen these two methods are consistent with each other. The outliers in the upper-right (Mgb) and lower-right (Fe5335) panels are caused by bad stacked spectra, i.e. our method is even more robust in calculating mean spectral indices for regions with a few problematic spaxels.

\begin{figure} 
\begin{center}
\includegraphics[scale=0.6]{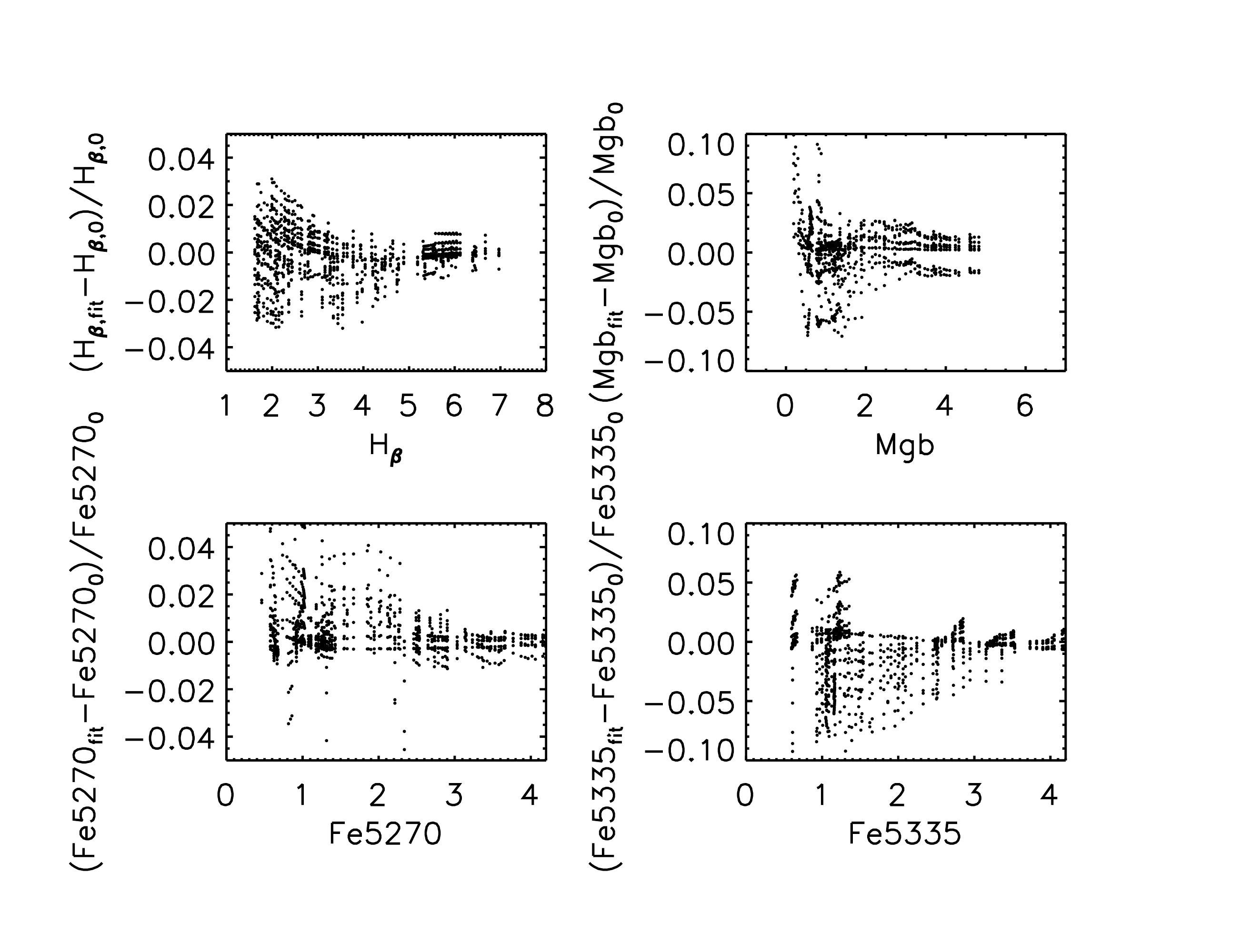}  
\caption{Comparing velocity dispersion corrected spectral indices using our method ($x_{fit}$) with spectral index values of the original \citet{Maraston2011} templates  ($x_0$, without velocity dispersion broadening). 
}
\label{vd_miles}
\end{center}
\end{figure}

Velocity dispersion broadening has a non-negligible effect on spectral index values, especially for spectra with stellar velocity dispersion values larger than 100 km/s. We therefore have to correct for velocity dispersion broadening in our spectral index measurements. Many previous studies perform velocity dispersion corrections in a very time consuming way: one fits the observed spectra with template spectra and then measures spectral indices using the fitted template spectra with and without velocity dispersion to derive the correction factor. Here we use a new method to correct for the velocity dispersion effect: we assume a simple polynomial relation between spectral indices with and without velocity dispersion broadening,
\begin{equation}
x_0 = p_0 + p_1 \,x^{p_2} + p_3 \,\sigma^{p_4} +p_5 \,x^{p_6}\,\sigma^{p_7},
\label{vd_fit}
\end{equation}
where $\sigma$ is velocity dispersion in unit of km/s,  $x$ and $x_0$ are the spectral index values before and after velocity dispersion correction, and $p_0 - p_7$ are fitting parameters. 
We use MILES template spectra \citep[which have similar spectral resolution to SDSS, ][]{sb06, Vazdekis et al.2010} with different velocity dispersion values to fit the  parameters in Eq. \ref{vd_fit}. The fitted parameters for each of the four spectral indices are listed in Table \ref{fitting_parameter}. 
We then verify the fitted parameters using the \citet{Maraston2011} single stellar population templates with \citet{chabrier03} IMF and MILES stellar library: 
we do velocity dispersion correction by applying Eq. \ref{vd_fit} with parameters in Table \ref{fitting_parameter} 
to spectral index values derived using template spectra with different velocity dispersions, 
and then compare the corrected values with spectral index values of the original \citet{Maraston2011} templates. 
The results are shown in Fig. \ref{vd_miles}. 
The corrected values show little difference from the real values (mostly within 2\%) except for small values ($<1$) of Mgb, Fe5270 and Fe5335, which are rarely seen in our galaxy sample. These parameters can also be applied to other observations which have a similar instrumental spectral resolution to SDSS.
Using these fitted parameters and measured velocity dispersion values, we are able to calculate spectral indices without velocity dispersion for little computational cost.  For our sample galaxies, we use velocity dispersion measurements (for each spaxel) from the MaNGA Data Analysis Pipline (DAP, Westfall et al. submitted), which uses the pPXF \citep{Cappellari2017} full spectral fitting technique in deriving kinematical properties.

\begin{table*}
\caption{Fitting parameters for Eq. \ref{vd_fit}. Note that the values of $p_5$ are very small, so we show $10^5\times p_5$ in this table. 
Also, $p_0$,$\,\,p_2$,$\,\,p_3$,$\,\,p_4$,$\,\,p_6$ and $p_7$ have been fixed at the specific values in the table in fitting for Mgb.}
\small\addtolength{\tabcolsep}{-2pt}
\begin{center}
\begin{tabular}{ccccccccc}
\hline
\hline
Name & $p_0$ & $p_1$ & $p_2$ & $p_3$ & $p_4$ & $10^5p_5$ & $p_6$ & $p_7$  \\
\hline
H$\beta$  &  0.15$\pm0.07$ &    0.92$\pm0.04$  &    1.03$\pm0.02$  & 0$\pm0.02$  &  0.57$\pm0.89$ & $0.11\pm0.17$  &    1.22$\pm0.21$ &    1.93$\pm0.21$    \\
\hline
Mgb    & 0 &  $0.83\pm0.15$ & 1 & 0 & 0 &  $110\pm50$  & 1 & 1   \\
\hline
Fe5270 & 0.54$\pm1.4$  &   0.97$\pm0.01$    &  1.01$\pm0.01$  &  -0.46$\pm1.39$ &  0.037$\pm0.10$ & $76\pm1.6$   &  0.87$\pm0.02$     & 1.48$\pm0.03$    \\  
Fe5335 &    -0.02$\pm0.02$    & 1.04$\pm0.01$  &  0.97$\pm0.01$ & 0$\pm0.008$  &   0.46$\pm0.4$ & $0.52\pm0.08$    &  1.01$\pm0.02$   &   2.03$\pm0.02$  \\
\hline
\hline
\label{fitting_parameter}
\end{tabular}
\end{center}
\end{table*}%

Star forming galaxies may have H$\beta$ emission, so we also have to correct for emission line flux when calculating the H$\beta$ index. We correct for  H$\beta$ emission by subtracting a Gaussian emission line profile with a total flux equal to the H$\beta$ emission flux value provided by the MaNGA DAP (Westfall et al. submitted).

\subsection{Effective stellar velocity dispersion}
\label{def_vd}
As shown in the introduction, stellar velocity dispersion is well correlated with the $\alpha$-to-iron ratio. 
Stellar velocity dispersion values from the literature \citep[e.g.][]{Liu et al.2016} are mostly measured using single fibers with fixed angular sizes 
or long slits with arbitrary apertures
and thus may introduce biases for very small or large galaxies. 
Here we use an effective stellar velocity dispersion $\sigma_*$ \citep{Li et al.2018, Cappellari et al.2013} estimated in a $0.3R_e$ aperture calculated for each galaxy. 
The $\sigma_*$ is defined as the square-root of the luminosity-weighted average second moments of the velocity within the  $0.3R_e$ aperture:
\begin{equation}
\sigma_*=\sqrt{\frac{\Sigma_k F_k (v^2_k+\sigma_k^2)}{\Sigma_kF_k}},
\end{equation}
where $v_k$ and $\sigma_k$ are the velocity and stellar velocity dispersion of the $k$-th IFU spaxel respectively, and $F_k$ is the flux in the $k$-th IFU spaxel. This $\sigma_*$ definition has been proven to be close (mostly within $\sim10\%$) to the velocity dispersion measured from a stacked spectrum using spectra within the same aperture \citep{Li et al.2018, Cappellari et al.2013}.

\subsection{Environment and formation history}

\begin{figure}
\begin{center}
    \includegraphics[scale=0.5]{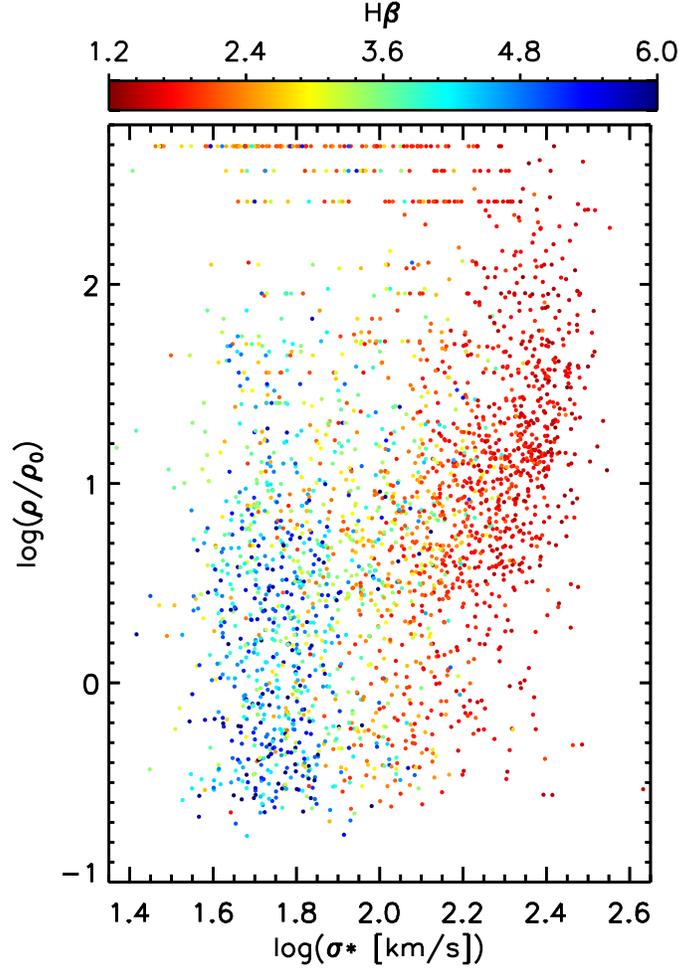}
    \caption{Galaxy local density ($\rho$) versus effective stellar velocity dispersion $\sigma_*$ diagram for our sample galaxies. Each dot is a galaxy, color-coded by the average H$\beta$ index value  within 0.3$R_e$ of each galaxy.  $\rho$ is in units of average cosmic mean density, $\rho_0=\rm 7.16\times10^{10}M_{\odot}/{\it h}/(Mpc\,{\it h}^{-1})^3$. }
    \label{fig:vd-den}
  \end{center}
\end{figure}

In this paper, we consider four environment indicators:   local mass density, types of large scale structure (LSS), isolated/central/satellite types, and projected halo-centric radius (for satellite galaxies only). 
The first two environment indicators are taken from the ELUCID project \citep[Exploring the Local Universe with the reConstructed Initial Density field;][]{Wang et al.2014}, which uses the halo-domain method developed by \citet{Wang et al.2009} to reconstruct the cosmic density field in the local universe from the SDSS DR7 galaxy group catalogue \citep{Yang et al.2007}. As shown in \citet{Wang et al.2016}, the reconstructed density field matches well the distributions of both the galaxies and groups. 
The local mass density environment indicator of a galaxy is the density at the position of the galaxy smoothed by a Gaussian kernel with a smoothing scale of 1 Mpc$/h$ (see Fig. \ref{fig:vd-den}, effective stellar velocity dispersion v.s. local density distribution for our sample galaxies).   
The morphology of the LSS is very complex and we adopt a dynamical classification method
developed by \citet{Hahn et al.2007}, which uses the eigenvalues of the tidal tensor to determine the type of the local 
structure in a cosmic web. The LSS environment is classified into four categories following the definition by \citet{Hahn et al.2007}: a $cluster$ has three positive eigenvalues (fixed points); a $filament$ has two positive and one negative eigenvalues (two-dimensional stable manifold); a $sheet$ has one positive and two negative eigenvalues (one-dimensional stable manifold); and a $void$ has three negative eigenvalues (unstable orbits). 
We refer the reader to \citet{Wang et al.2016} for details of the reconstruction methods and \citet{Zheng et al.2017} for more examples of utilizing these two environment indicators.  
Each galaxy is also classified into three types according to \citet{Yang et al.2007}: isolated (the only galaxy member in a group), central (the most massive galaxy in a group) and satellite. For satellite galaxies, we also examine dependences on their projected halo-centric radii \citep{Yang et al.2007}.

The ELUCID project also derives initial conditions for our universe using the reconstructed density field in the SDSS DR7 region and use the initial conditions to simulate the formation history of dark matter halos in this region. This ELUCID simulation can reliably reproduce most of the massive groups \citep{Wang et al.2016, Wang et al.2018}. We use the formation redshift derived from the ELUCID simulation to indicate galaxy formation histories and investigate dependance of Mgb/$\langle$Fe$\rangle$ on  formation redshift in the following sections. The formation redshift, $z_f$,  is defined as the highest redshift at which half of the final halo mass has assembled into progenitors more massive than $10^{11.5}h^{-1}M_{\odot}$ \citep{Wang et al.2018}. 

Note that not all galaxies in our sample have environment information. The numbers of galaxies with different environment measurements are shown in Table \ref{envtable}. 

\begin{table*}
\caption{Numbers of galaxies with different environment measurements. }
\begin{center}
\begin{tabular}{lc}
\hline
\hline
Type & Number of galaxies \\
\hline
All &  2997 \\
\hline
Local density & 2447 \\
LSS type & 2447 \\
Isolated/Central/Satellite & 2663 \\
Projected halo-centric radius (satellite only) & 811 \\
Formation redshift (all types) &  2423 \\
Formation redshift (central\,\&\,isolated only) &  1632 \\

\hline
\hline
\label{envtable}
\end{tabular}
\end{center}
\end{table*}%

\section{Results}
\label{results}
\subsection{Mgb/$\langle$Fe$\rangle$ - $\sigma_*$ relation}

\label{mgb2fe-sigma}
The spectral index ratio Mgb/$\langle$Fe$\rangle$, where
$\rm \langle Fe \rangle = (Fe5270+Fe5335)/2$,  is a widely used indicator in previous studies for the $\alpha$-to-iron ratio  \citep[see Fig. 4 of][]{Thomas et al.2003}.  
We note that $\rm Mgb/\langle Fe \rangle$ is indicative for general element ratio trends but detailed element ratio calculations would require 
more sophisticated techniques such as model fitting (see discussion in Section \ref{discussion_implications}). 
We also derived $\alpha$-to-iron ratio ([$\alpha$/Fe]) by fitting the spectral indices with \citet{Thomas et al.2011} models, however, the relations and environmental dependence of [$\alpha$/Fe] are less obvious because of large scatters. We therefore focus on Mgb/$\langle$Fe$\rangle$ in the main text and append the [$\alpha$/Fe] related results in the Appendix \ref{appendix}.

\begin{figure} 
\begin{center}
\includegraphics[scale=0.5]{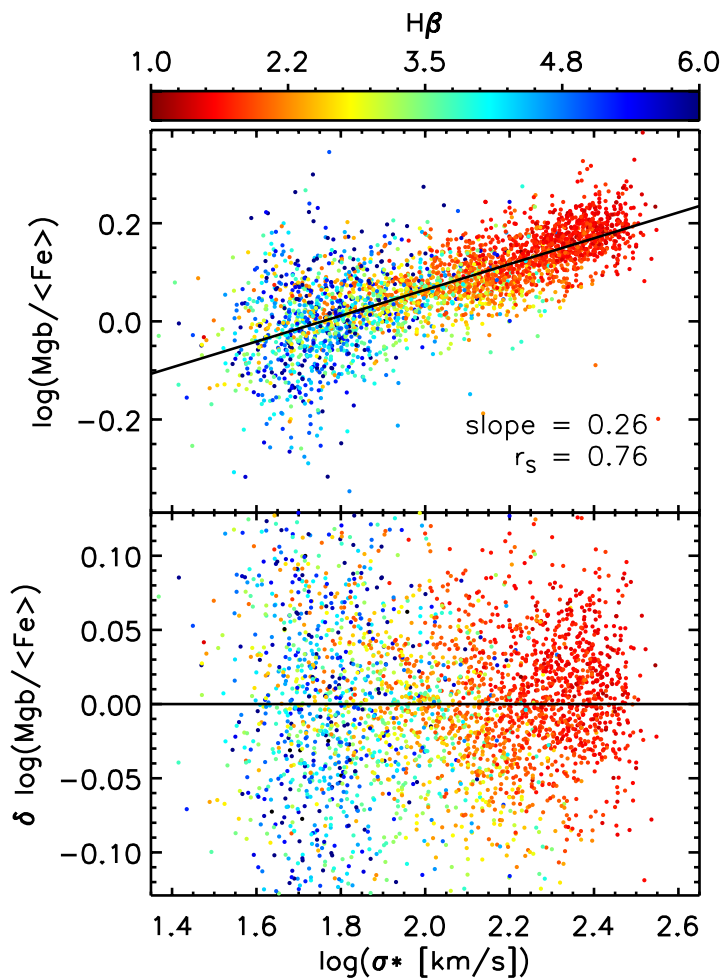}  
\caption{Upper panel: Mgb/$\langle$Fe$\rangle$ versus velocity dispersion for the central regions (within $\rm R_e/3$); the solid line is a linear fit to all the data points. The slope of of the fitted line and the Spearman's rank correlation coefficient ($r_S$) of the data are shown on the plot. Lower panel: the residuals, i.e. deviations of individual data points of the upper panel from the fitted straight line. Symbols are color-coded by H$\beta$. }
\label{sindx_sigma_bulge_den}
\end{center}
\end{figure}

It has been claimed that Mgb/$\langle$Fe$\rangle$ (or [$\alpha$ /Fe]) is best correlated with stellar velocity dispersion \citep[e.g.][]{Kuntschner et al.2010, Thomas et al.2010, Liu et al.2016}. We hence plot Mgb/$\langle$Fe$\rangle$ versus the effective stellar velocity dispersion $\sigma_*$  for the central regions (within 0.3$R_e$) of all our sample galaxies in Fig. \ref{sindx_sigma_bulge_den}.

It is obvious that Mgb/$\langle$Fe$\rangle$ is well correlated with $\sigma_*$ and most galaxies lie in close vicinity around the linearly fitted straight line (solid line in the upper-panel of Fig. \ref{sindx_sigma_bulge_den}). The Spearman's rank correlation coefficient is $r_S=0.76$. This means that not only early type but all types of galaxies in our sample follow the Mgb/$\langle$Fe$\rangle$-$\sigma_*$ relation.  We can also see from the color coding that high-$\sigma_*$ galaxies mostly have small H$\beta$ values (relatively old) and low-$\sigma_*$ galaxies mostly have large H$\beta$ values (relatively young).  Low-$\sigma_*$ ($<80\rm\,km/s$) galaxies have larger scatters but are still centered around the solid line, which is consistent with \citet{Liu et al.2016}.

\subsection{Environmental dependence}

\begin{figure} 
\begin{center}
\includegraphics[angle=90,scale=0.6]{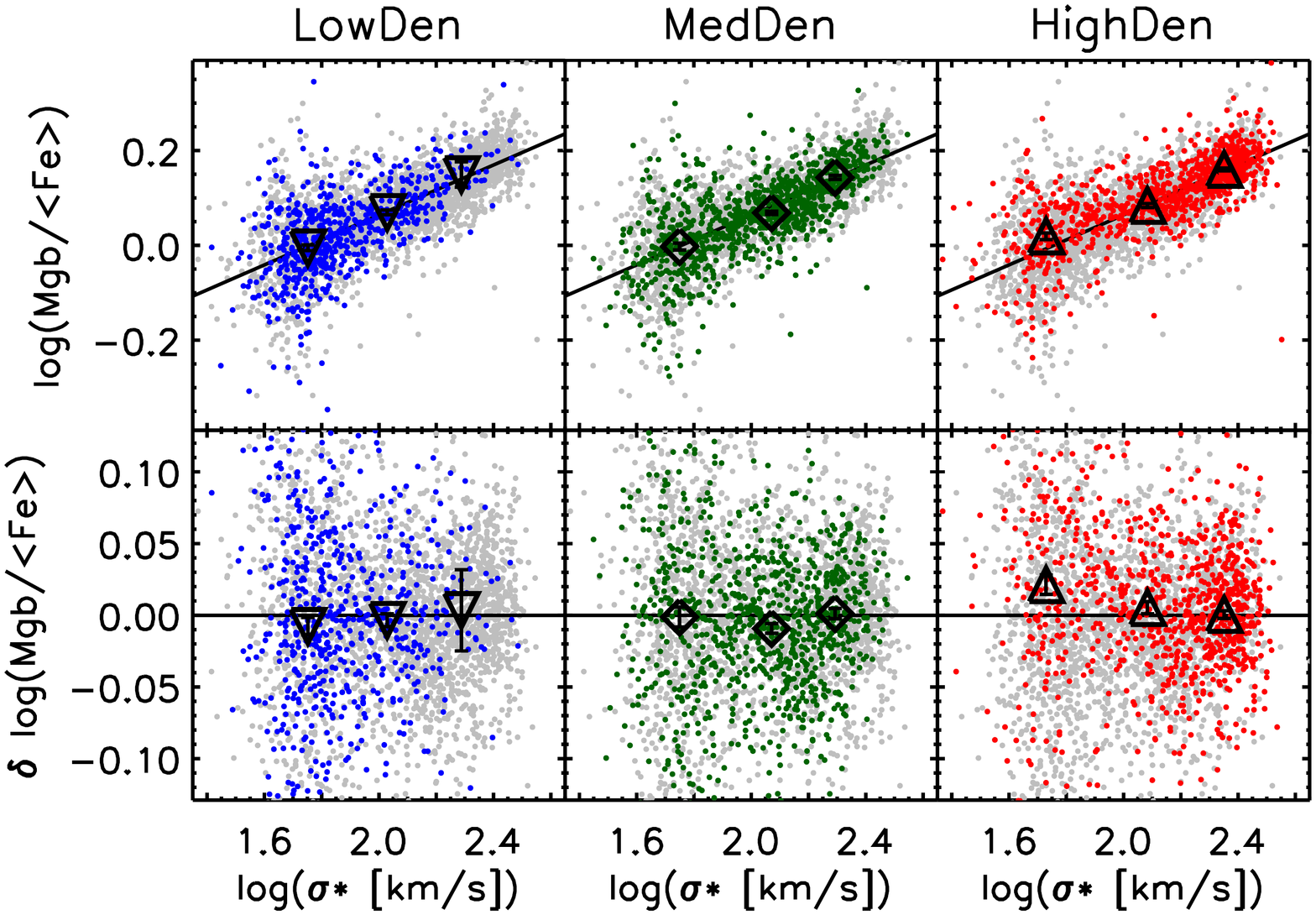}  
\caption{
$\rm Mgb/\langle Fe\rangle$ -$\sigma_*$ relation in different local density environments.
This plot is similar to Fig. \ref{sindx_sigma_bulge_den} but separated into three different local density environments: low-density ($\rho<2\rho_0$, left panels), medium-density ($2\rho_0<\rho<12.6\rho_0$, middle panels) and high-density ($\rho>12.6\rho_0$, right panels), where $\rho$ is the average density within 1 Mpc of each galaxy and $\rho_0$ is the average cosmic mean density, $\rm 7.16\times10^{10}M_{\odot}/{\it h}/(Mpc\,{\it h}^{-1})^3$. In each panel, gray points in the background show the whole sample and the  blue, green and red points show galaxies in low, medium and high density regions respectively.  The big symbols are median values of the colored points in three $\sigma_*$ bins: low-$\sigma_*$ ($\sigma_*<80\,\rm km/s$), medium-$\sigma_*$ ($80\, \rm km/s <\sigma_* <158\, \rm km/s$) and high-$\sigma_*$ ($\sigma_*>158\,\rm km/s$). The solid lines are the same as in Fig. \ref{sindx_sigma_bulge_den}. Note the lower-panels have a smaller y-axis range.}
\label{sindx_den_bottom_bulge}
\end{center}
\end{figure}

As shown in the previous subsection, $\rm Mgb/\langle Fe\rangle$ is well correlated with $\sigma_*$. Therefore, in order to examine the environmental dependence, we need to take out the effect by $\sigma_*$.
We firstly follow the method of \citet{Liu et al.2016} and divide galaxies into three different velocity dispersion ranges: low-$\sigma_*$ ($\sigma_*<80\,\rm km/s$), medium-$\sigma_*$ ($80\, \rm km/s <\sigma_* <158\, \rm km/s$) and high-$\sigma_*$ ($\sigma_*>158\,\rm km/s$). 
We further divide galaxies in these $\sigma_*$ ranges into three different environments based on local densities: low-density ($<2\rho_0$), medium-density ($2\rho_0<\rho<12.6\rho_0$) and high-density ($>12.6\rho_0$), where $\rho_0$ is the average cosmic mean density, $\rm 7.16\times10^{10}M_{\odot}/{\it h}/(Mpc\,{\it h}^{-1})^3$.  
The velocity and local density ranges are chosen so that each velocity or local density range has roughly the same number of galaxies.
Fig. \ref{sindx_den_bottom_bulge} shows the $\rm Mgb/\langle Fe\rangle$ -$\sigma_*$ relation for the central regions (within $\rm R_e/3$) in the three different local density ranges. In the lower panels of  Fig. \ref{sindx_den_bottom_bulge}, we plot the residuals ($\rm \delta \log (Mgb/\langle Fe \rangle$)). We also plot average values of galaxies in these three different environments and three different $\sigma_*$ ranges. Low- and medium-$\sigma_*$ galaxies in the high-density environment (upward-triangle) show enhancement in Mgb/$\langle$Fe$\rangle$.

\begin{figure} 
\begin{center}
\includegraphics[scale=0.6]{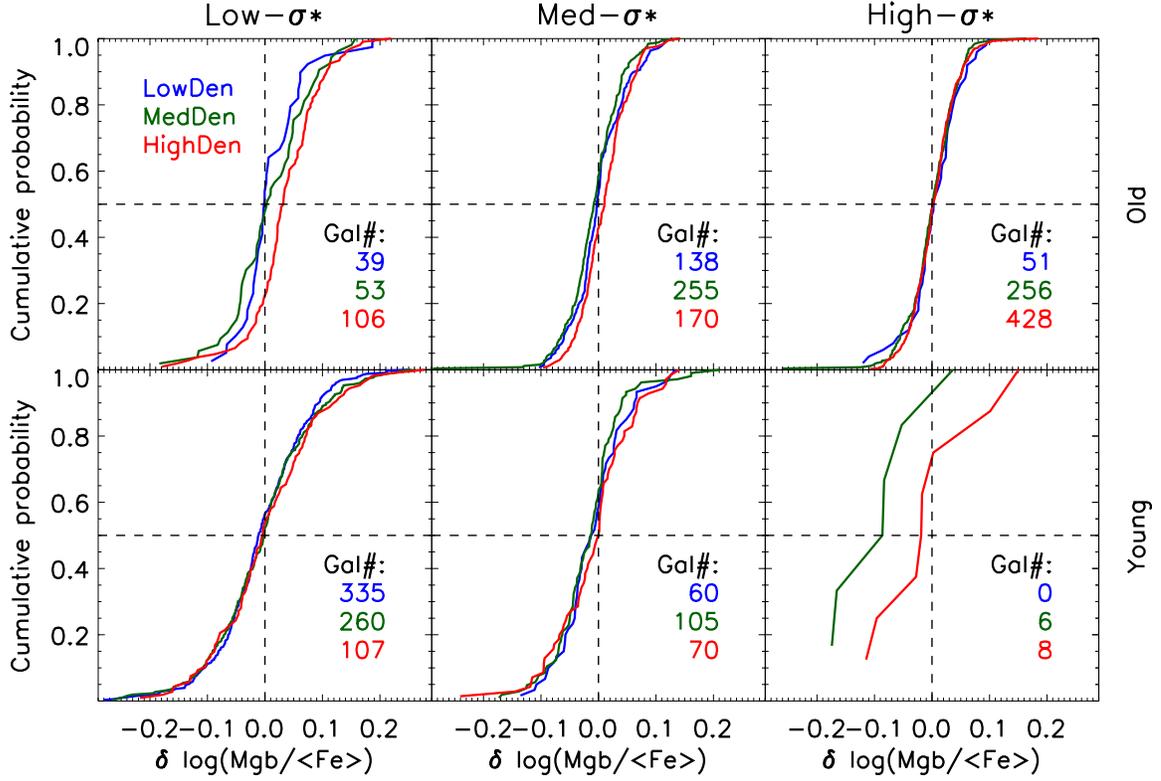}  
\caption{Cumulative probability distribution  of the residuals ($\rm \delta \log (Mgb/\langle Fe \rangle$)) for central regions of galaxies in different H$\beta$, $\sigma_*$ and local density environment bins. The upper-panels show old (central H$\beta<3$) galaxies and lower-panels show young (central H$\beta>3$) galaxies. The left, middle and right panels show low-$\sigma_*$ ($\sigma_*<80$km/s), medium-$\sigma_*$ ($\rm 80km/s<\sigma_*<158km/s$), and high-$\sigma_*$ ($\sigma_*>158$km/s)  galaxies respectively. In each panel, red, green and blue lines show galaxies in low density, medium density and high density environments respectively. The numbers show the number of galaxies in their corresponding H$\beta$-$\sigma_*$-local density environment bins. }
\label{sindx_den_bottom_hist}
\end{center}
\end{figure}

\citet{Thomas et al.2010} showed that old and young galaxies may have different dependance on environments. 
We therefore also divide our sample into `old' (central H$\beta<3$) and `young' (central H$\beta>3$) populations using the H$\beta$ index and plot cumulative probability distributions in Fig. \ref{sindx_den_bottom_hist}  of the residuals for the central regions of galaxies in different H$\beta$, $\sigma_*$ and local density environment bins. 
For `old' galaxies (upper panels), we see a clear large residual in the low-$\sigma_*$ high-local density bin (red line in upper-left panel). 
The red line (high-local density) in the medium-$\sigma_*$ bin (upper-middle panel) is also large comparing to other lines (lower local densities) although less obvious.
The difference  disappears in the high-$\sigma_*$ bin (upper-right panel).  
For `young' galaxies (lower panels),  the environmental dependance is not clear.

\begin{figure} 
\begin{center}
\includegraphics[scale=0.6]{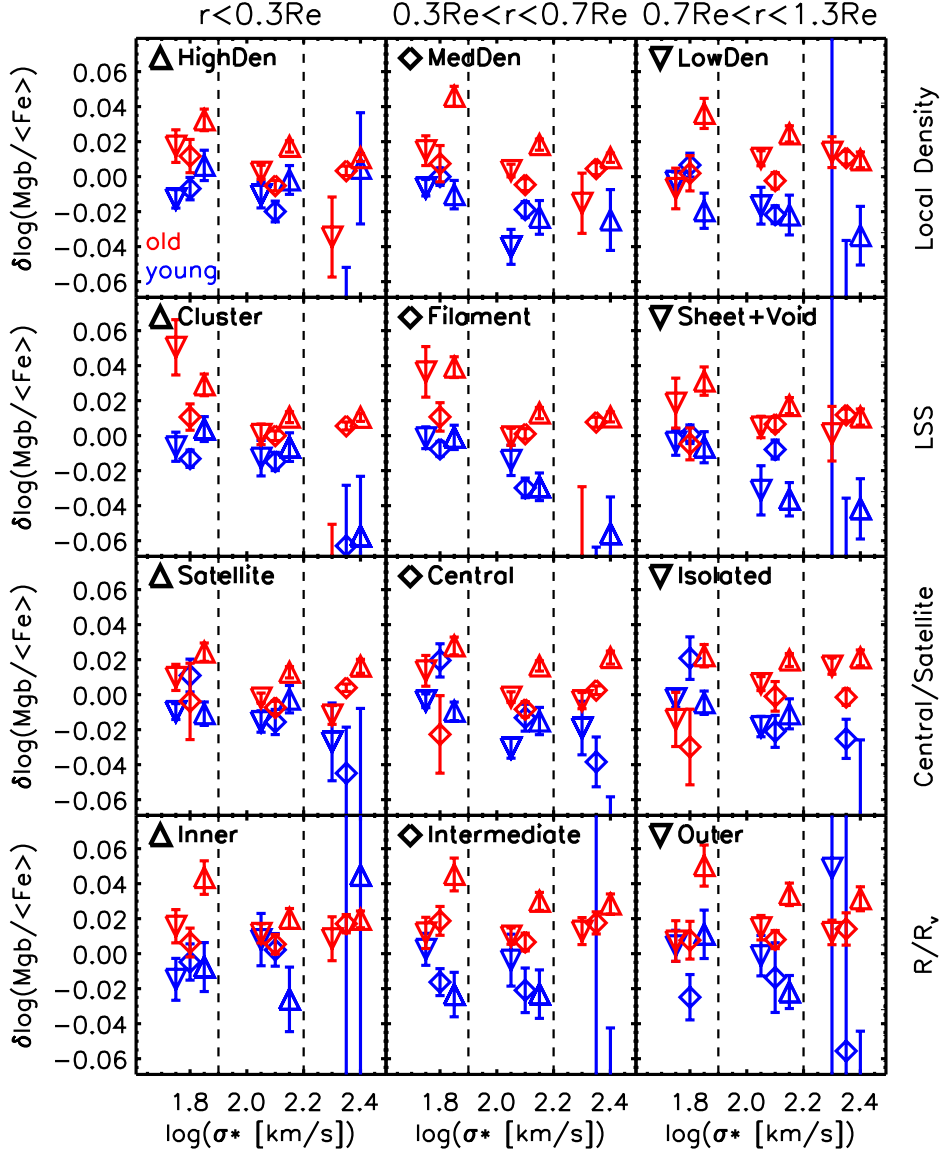}  
\caption{
\footnotesize
Weighted mean values of the residuals ($\rm \delta \log (Mgb/\langle Fe \rangle$)) for different galaxy regions and different galaxy environment indicators. 
The columns are ordered in galaxy regions: central ($r<0.3R_e$, left column), intermediate ($0.3R_e<r<0.7R_e$, middle column), and outer ($0.7R_e<r<1.3R_e$, right column).
The rows are ordered in environment indicators: local density (top row), large scale structure (second row), isolated/central/satellite (third row) and radii to group centers (bottom row, satellite galaxies only). 
The upward triangles, downward triangles and diamonds for each row are denoted in the legends of the plot. 
Red symbols are for galaxies with H$\beta<3$ (old) and blue symbols are for galaxies with H$\beta>3$ (young).
The plots are the same as the lower panels of Fig. \ref{sindx_den_bottom_bulge} but with a narrower y-axis range. 
}
\label{sindx_env}
\end{center}
\end{figure}

Since we have IFU data for all our sample galaxies, we can investigate the environmental dependence of the residuals for the intermediate and outer regions in addition to the central regions of galaxies. We calculate weighted mean values of the residuals in bins constructed with different galaxy regions, H$\beta$, $\sigma_*$ and local density environments 
and plot them in the first row of Fig. \ref{sindx_env}. The weights used in this calculation are from \citet{Wake et al.2017}. For all three regions,  `old' low-$\sigma_*$  galaxies in a high local density environment show an obvious large residual, however, this kind of signature is not obvious for `young' galaxies and/or high-$\sigma_*$ galaxies. 

We also explore dependence on LSS types, isolated/central/satellite types and make similar plots using these environment indicators  in the second and third rows of Fig. \ref{sindx_env}.  There is no clear systematic trend along LSS types (second row),
and similarly there is no obvious systematic trend along isolated/central/satellite types either (third row).  
`Old' satellite galaxies (in the third row), however, have the highest residual in all $\sigma_*$ bins and all galaxy regions.  

For satellite galaxies, we divide them into three groups according to their projected halo-centric radii (in units of virial radius of the group $R_V$): inner ($R<0.18R_V$), intermediate ($0.18R_V<R<0.7R_V$) and outer ($R>0.7R_V$); and make plots accordingly in the bottom row of Fig. \ref{sindx_env}. The residuals for `old' low-$\sigma_*$  galaxies in the inner regions ($R<0.18R_V$) of galaxy groups are obviously large comparing to those of galaxies in the  outer and intermediate regions. `Old' medium-$\sigma_*$ and high-$\sigma_*$ galaxies also show a similar trend but less obviously. There is no clear systematic trend for `young' galaxies. Note that the number of satellite galaxies is relatively small (cf. Table \ref{envtable}).

\subsection{Formation redshift dependence}

\begin{figure} 
\begin{center}
\includegraphics[scale=0.6]{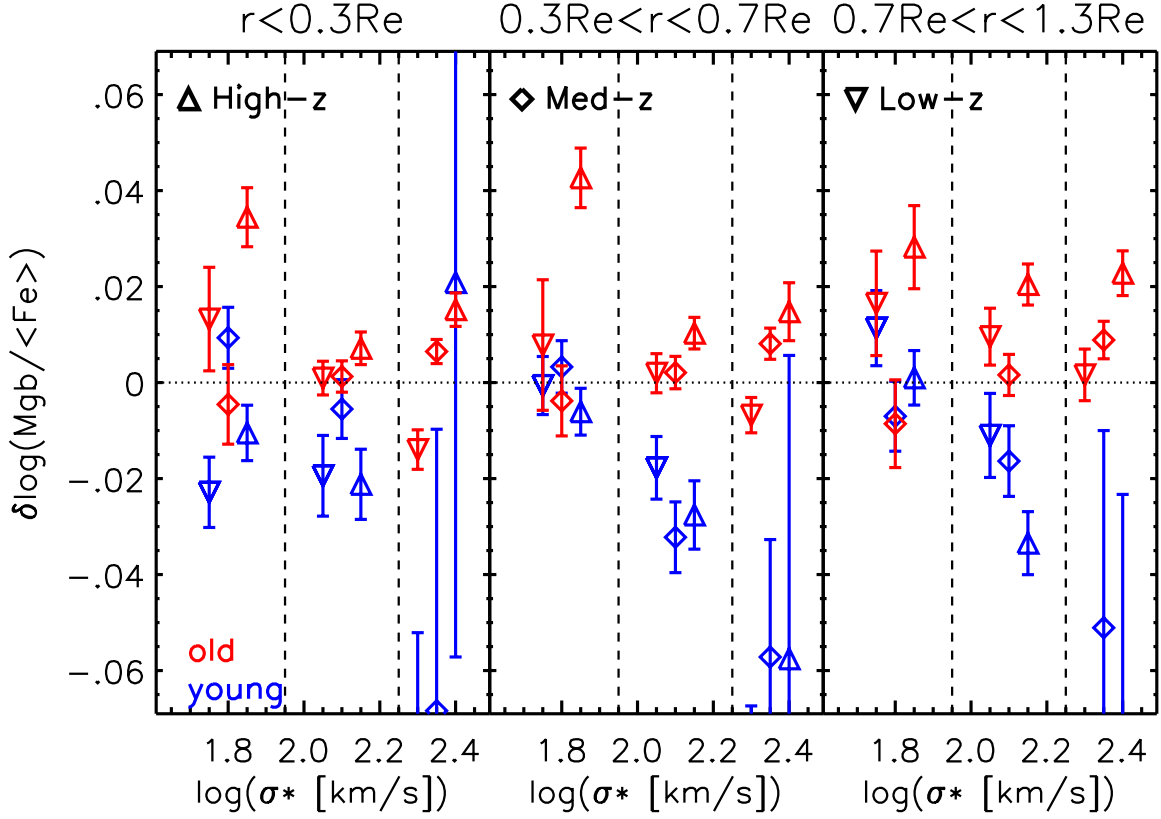}  
\caption{
\footnotesize
Weighted mean values of the residuals ($\rm \delta \log (Mgb/\langle Fe \rangle$)) for  different parts of galaxies: central ($r<0.3R_e$, left panel), intermediate ($0.3R_e<r<0.7R_e$, central panel), and outer ($0.7R_e<r<1.3R_e$, right panel) with different dark matter halo formation histories. The plot is the same as the Fig. \ref{sindx_env} but only shows dependence on the formation redshift: low-z ($z_f<0.9$, upward triangles), medium-z ($0.9<z_f<1.5$, downward triangles) and high-z ($z_f>1.5$, diamonds). Red symbols are galaxies with H$\beta<3$ (old) and blue symbols are galaxies with H$\beta>3$ (young).}
\label{sindx_zf_all}
\end{center}
\end{figure}

We investigate dependence on galaxy halo formation redshift ($z_f$) and the results are shown in Fig. \ref{sindx_zf_all}. 
We do not see any clear systematic trend along formation redshifts, however, for `old' galaxies, those with high formation redshift (formed earlier) have relatively larger residuals. This is true for almost all $\sigma_*$ bins and all galaxy regions. For `young' galaxies, this signature is not obvious.

\begin{figure} 
\begin{center}
\includegraphics[scale=0.6]{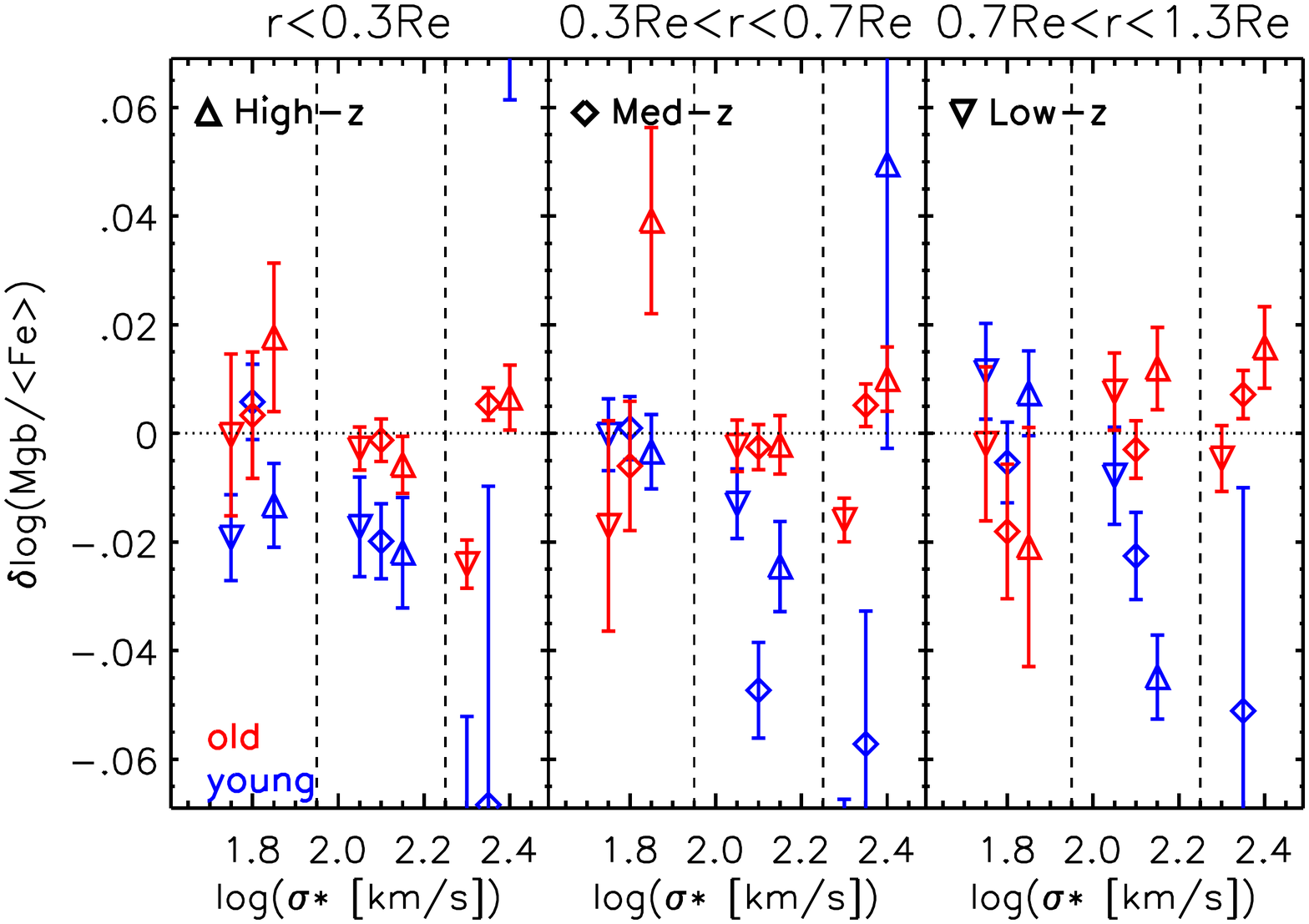}  
\caption{
\footnotesize
This plot is similar to Fig. \ref{sindx_zf_all} but using isolated and central galaxies only. }
\label{sindx_zf_ceniso}
\end{center}
\end{figure}

According to \citet{Wang et al.2018}, the $z_f$ provided is the formation redshift of the main halo and mostly reflects the formation time of the central/isolated galaxy. Satellite galaxies residing in separate sub-halos may have formation times different from the main halo. Consequently, we remove satellite galaxies and remake the plot showing dependance on $z_f$ as Fig. \ref{sindx_zf_ceniso}.  Generally, error bars are larger by comparison with Fig. \ref{sindx_zf_all}.  For `old' galaxies, those with high formation redshift (formed earlier) still have relatively larger residuals but not in all panels. For `young' galaxies, there is still no clear signatures.

\subsection{Gradients of Mgb/$\langle$Fe$\rangle$ within galaxies}

\begin{figure} 
\begin{center}
\includegraphics[scale=0.38]{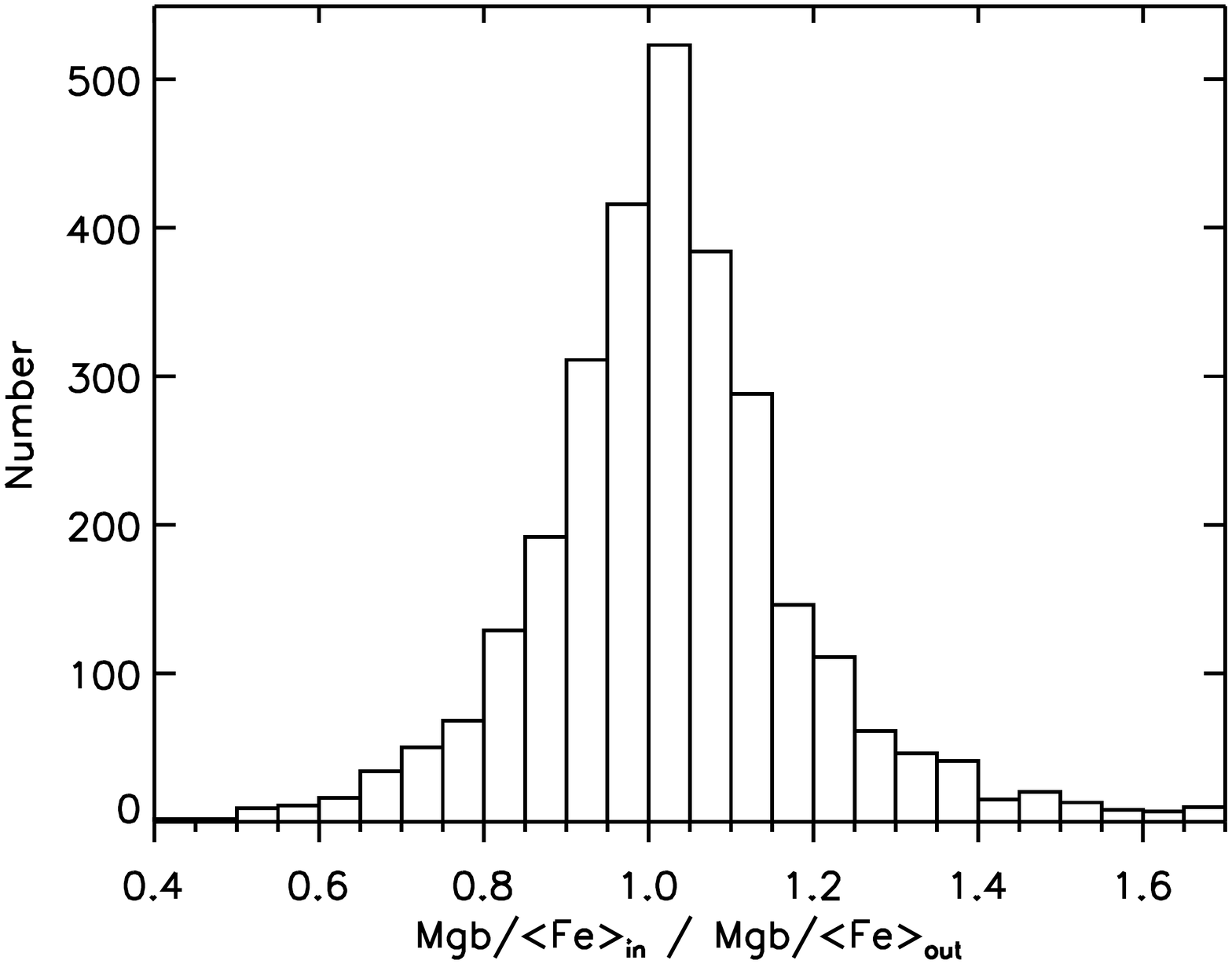}  
\caption{Distribution of the inner-to-outer ratio of Mgb/$\langle$Fe$\rangle$ for our sample galaxies. The inner and outer regions are defined as $r<0.3R_e$ and $0.7R_e<r<1.3R_e$ respectively. }
\label{sindx_inout}
\end{center}
\end{figure}

The other interesting study enabled by IFU observations is Mgb/$\langle$Fe$\rangle$ gradients within galaxies. However, we only have three radial bins for each galaxy because of our requirement for high SNR. We therefore study this question in a simple way: plot histograms of the inner-to-outer ratio of Mgb/$\langle$Fe$\rangle$ (see Fig. \ref{sindx_inout}). The inner and outer regions are defined as $r<0.3R_e$ and $0.7R_e<r<1.3R_e$ respectively. The average gradient for all galaxies is flat (inner-to-outer ratio of Mgb/$\langle$Fe$\rangle$ close to 1). 
We tried to divide galaxies into different environments, but still no gradients are observed for galaxies in different environments. 

\begin{figure} 
\begin{center}
\includegraphics[scale=0.4]{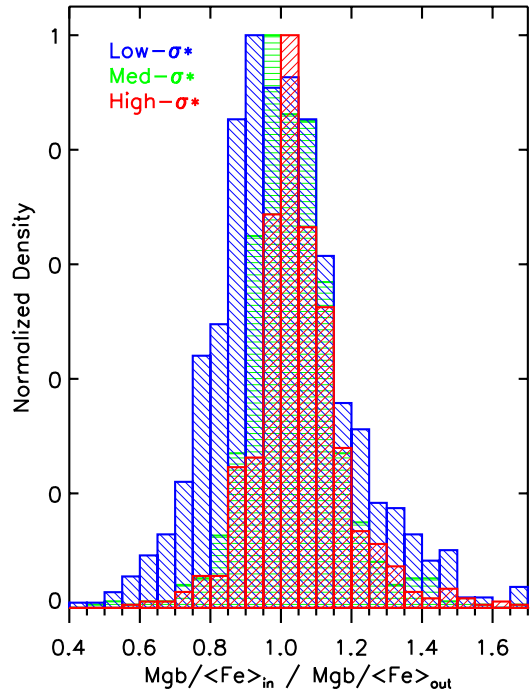}
\caption{Distributions of the inner-to-outer ratio of Mgb/$\langle$Fe$\rangle$ in different $\sigma_*$ bins. Histograms are color-coded by $\sigma_*$. }
\label{sindx_inout_vd}
\end{center}
\end{figure}

We then separate galaxies into different $\sigma_*$ bins in Fig. \ref{sindx_inout_vd}. This shows a more obvious difference: galaxies with high-$\sigma_*$ have a peak value around 1 (flat, zero-gradient) but with an extended tail towards larger values (negative gradient), while galaxies with low-$\sigma_*$ tend to have a positive Mgb/$\langle$Fe$\rangle$  gradient.

\begin{figure} 
\begin{center}
\includegraphics[scale=0.6]{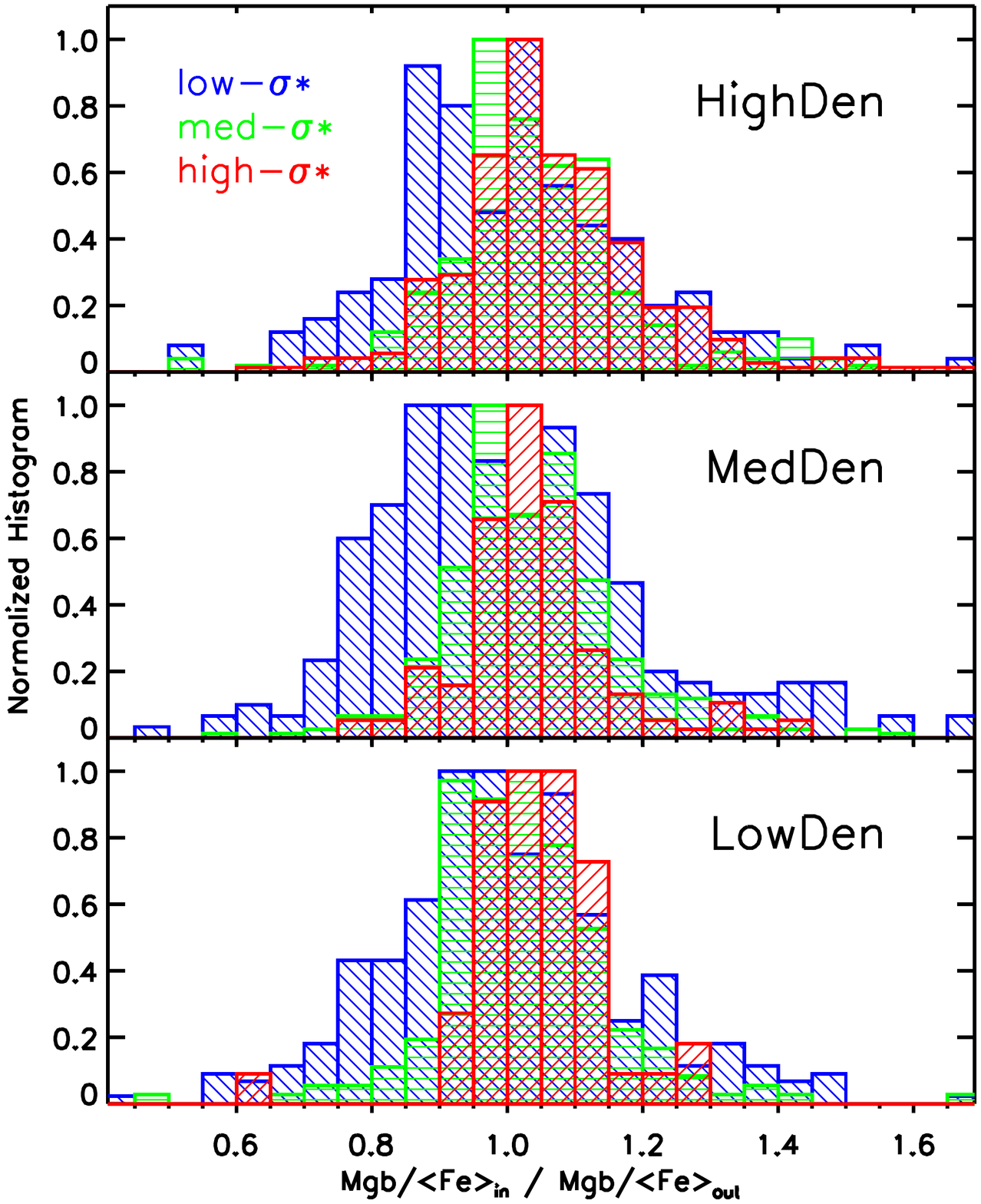}
\caption{Distributions of the inner-to-outer ratio of Mgb/$\langle$Fe$\rangle$ in different $\sigma_*$ and local density bins. The upper, middle and lower panels show galaxies in high local density, medium local density and low local density environments respectively. The histograms are color-coded by $\sigma_*$.  }
\label{sindx_inout_vdden}
\end{center}
\end{figure}

\begin{figure} 
\begin{center}
\includegraphics[scale=0.6]{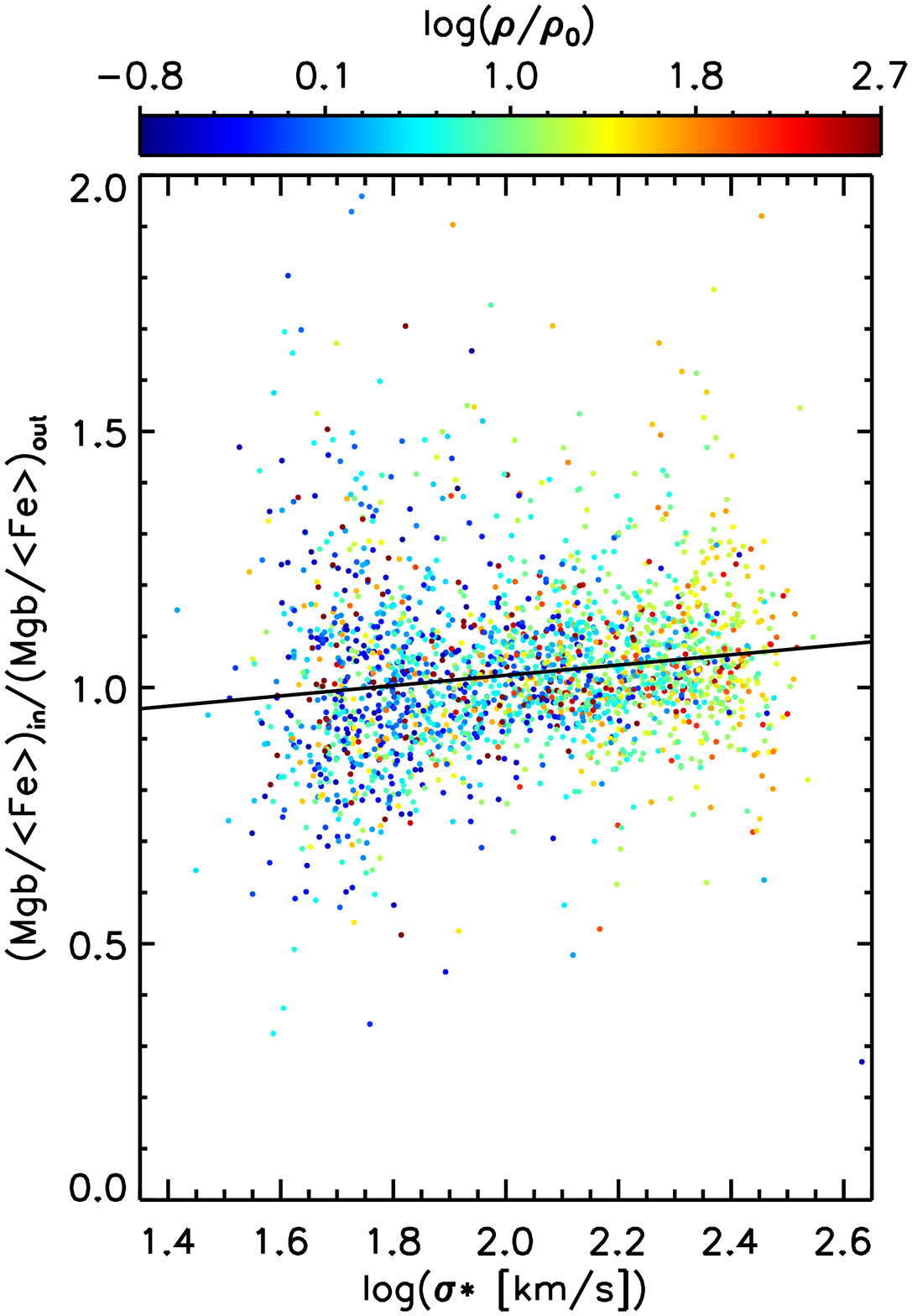}
\caption{Inner-to-outer ratio of Mgb/$\langle$Fe$\rangle$ versus $\sigma_*$. Each dot show one galaxy and the dots are color-coded by local density $\rho/\rho_0$. The straight line is a linear fit to all the dots.}
\label{sindx_inout_vdden2}
\end{center}
\end{figure}

We further divide galaxies into different $\sigma_*$-environment bins, but do not see a clear dependence on local density either. One example using $\sigma_*$-local density bins is shown in Fig. \ref{sindx_inout_vdden}. We also plot the inner-to-outer ratio of Mgb/$\langle$Fe$\rangle$ versus $\sigma_*$ color-coded by local density in Fig. \ref{sindx_inout_vdden2}. We can see a clear dependence of Mgb/$\langle$Fe$\rangle$ gradients on $\sigma_*$ and we see no clear dependence on local density.

\section{Discussion}
\label{discussion}

\subsection{Comparison with other studies}
\begin{figure} 
\begin{center}
\includegraphics[scale=0.6]{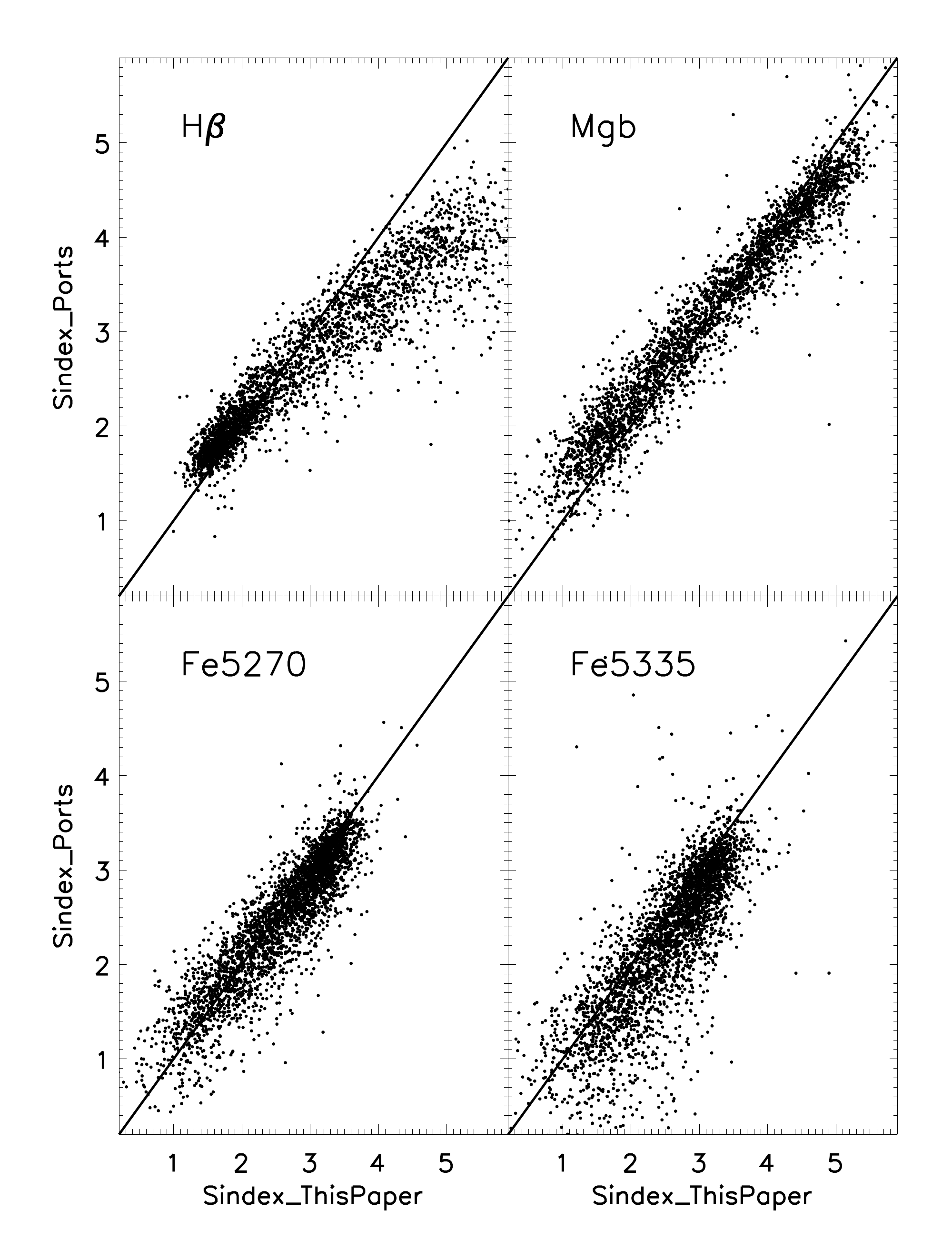}  
\caption{The four panels show the comparison of our (x-axes) and the Portsmouth VAC (y-axes) measurements of the four Lick indices (H$\beta$, Mgb, Fe5270 and Fe5335) using the whole MaNGA MPL-7 galaxy sample. Each dot shows the average Lick index value of spaxels within the central $6\arcsec$ aperture of an individual galaxy. The solid diagonal lines indicate index equality.}
\label{sindx_compare}
\end{center}
\end{figure}

We compare in Fig. \ref{sindx_compare} our spectral index measurements with the Portsmouth VAC (MaNGA Value Added Catalog produced by Portsmouth University).  Our measurements are consistent with the Portsmouth VAC, which uses the same sample but with a different algorithm. 
The spectral index measurement algorithm used by the Portsmouth VAC is similar to the MaNGA DAP code (Westfall et al. submitted) and the algorithm of  \citet{Parikh et al.2018}. We note that there are some discrepancies for high H$\beta$ ($>3$), however, we only use H$\beta$ values to classify `old' and `young' galaxies and it does not cause much discrepancies in this classification.

\begin{figure} 
\begin{center}
\includegraphics[scale=0.6, angle=90]{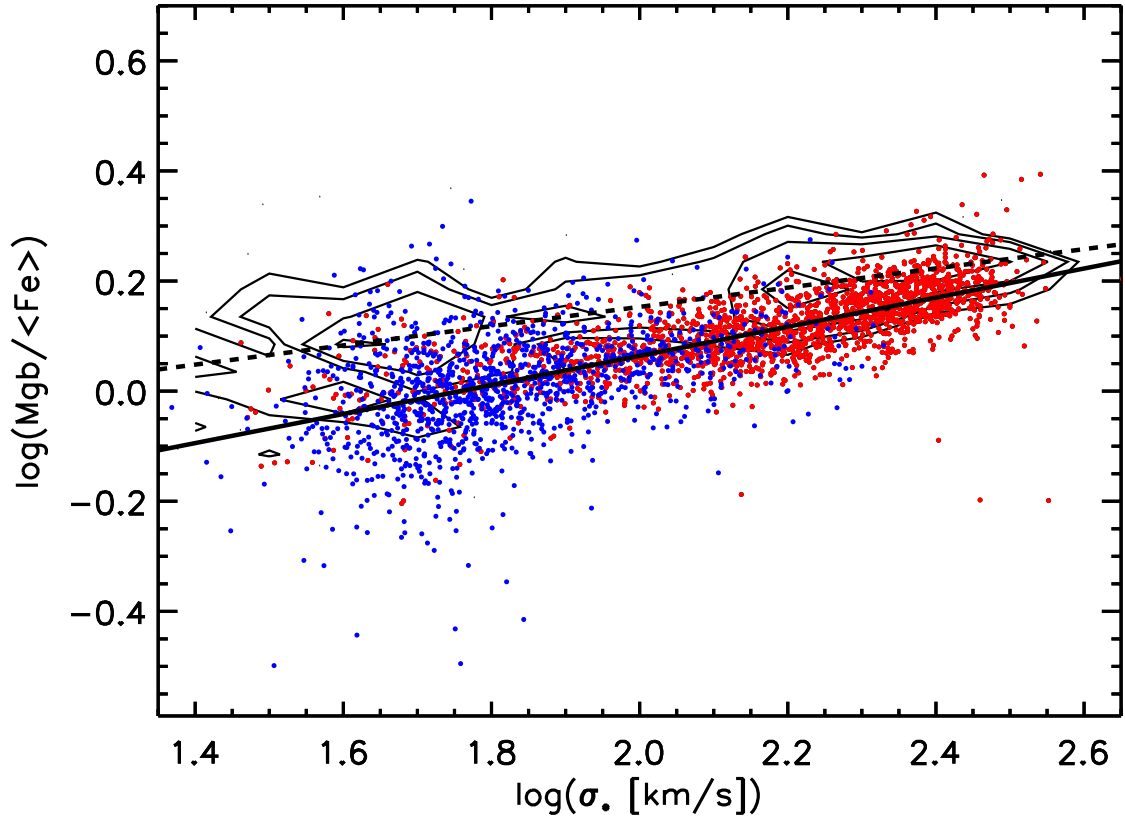}  
\caption{Comparison of Mgb/$\langle$Fe$\rangle$ between archival data compiled by \citet{Liu et al.2016} and this work. The contours are archival data from \citet{Liu et al.2016} and the points are our measurements for central regions of our sample galaxies. The red symbols are for galaxies with H$\beta<3$ (old) and the blue symbols are for galaxies with H$\beta>3$ (young). Note that the \citet{Liu et al.2016} data are collected from literatures, including data from \citet{Thomas et al.2010}. They only include early type galaxies and have different definition of $\sigma_*$ from this paper.}
\label{sindx_compare_liu}
\end{center}
\end{figure}

A comparison between our measurements of Mgb/$\langle$Fe$\rangle$ and various archival data from \citet{Liu et al.2016} is shown in Fig. \ref{sindx_compare_liu}. 
Our Mgb/$\langle$Fe$\rangle$-$\sigma_*$ relation is steeper than that of \citet{Liu et al.2016} and has a $\sim 0.05$ dex systematic difference at the high-mass end.
Note that data from different sources have different calibration procedures;  our spectral index values are corrected to the SDSS resolution ($\rm \sim 70km/s$) while data collected in \citet{Liu et al.2016} may have different spectral resolutions; also \citet{Liu et al.2016} has different aperture definitions from us when measuring spectral indices and velocity dispersions. 

\subsection{Implications for galaxy formation histories}
\label{discussion_implications}
Our results show that both `old' and `young' galaxies have good correlations between Mgb/$\langle$Fe$\rangle$ and $\sigma_*$, which is consistent with previous studies. 
The dependence of the residuals on local density environment (as well as projected halo-centric radius) is slight but significant enough to be identified visually. Also, the environmental dependence seems to be more obvious in low- and medium-$\sigma_*$ galaxies than in high-$\sigma_*$ galaxies. This implies that $\sigma_*$ or galaxy mass is the main parameter driving $\alpha$ enhancement and that environments only have  significant effects for low- and medium-mass galaxies. Low-mass galaxies in dense regions may have experienced fast quenching due to environmental effects and thus have a shorter star formation time scale resulting in a higher $\alpha$-to-iron ratio or Mgb/$\langle$Fe$\rangle$. 
`Young' galaxies generally have lower residuals in comparison to `old' ones,
and they do not show clear environmental dependence. This might be explained by the
fact that  the `young' galaxies have recent star formation.  The young stars in these
galaxies are expected to have formed from  the ISM that is polluted  by Type Ia
supernovae that have a  lower $\alpha$-to-iron ratio. This recent star formation could
also reduce the fast quenching signature produced by the environment.

The gradients of Mgb/$\langle$Fe$\rangle$ are close to 0 (flat), which is consistent with recent spatially resolved spectroscopic studies \citep[e.g.][]{Mehlert et al.2003, Kuntschner et al.2010, Spolaor et al.2010, Greene et al.2015, McDermid et al.2015, Parikh et al.2019}.
The gradients are less effected by environment but more determined by $\sigma_*$ or galaxy mass (see Fig. \ref{sindx_inout_vd}). 
\citet[][Fig. 14]{Kuntschner et al.2010} using 48 early type galaxies showed that their [$\alpha$/Fe] gradients are slightly negatively correlated with stellar velocity dispersion but does not depend on environment. 
This is also consistent with previous studies on environmental dependence of stellar population gradients such as \citet{Zheng et al.2017} and \citet{Goddard et al.2017}. 
This implies that the internal stellar population structures of galaxies are more affected by secular evolution processes driven by parameters such as galaxy mass. 

We note that the spectral indices Mgb, Fe5270 and Fe5335 do not only depend on chemical abundances, but also on age, metallicity and star formation histories \citep{Maraston et al.2003, Thomas et al.2003}. The index ratio Mgb/$\langle$Fe$\rangle$ does show a clear positive relation with [$\alpha$/Fe] but with larger scatters with larger age range \citep{Thomas et al.2003}. Our separation using H$\beta$ has partly removed some of the degeneracy introduced by age.

\section{Summary and conclusion}
\label{summary}

We derived H$\beta$, Mgb, Fe5270 and Fe5335 index maps for all MaNGA MPL-7 (equivalent to SDSS DR15) galaxies. We studied $\sim3000$ MaNGA galaxies and use Mgb/$\langle$Fe$\rangle$ as an $\alpha$-to-iron ratio indicator to study $\alpha$-abundance distribution within these galaxies. We studied environmental dependence of Mgb/$\langle$Fe$\rangle$-$\sigma_*$ relation using different different environment indicators: local density, large scale structure type, isolated/central/satellite type, and projected halo-centric radius. We also investigated the effects of galaxy halo formation redshift as well as gradients of Mgb/$\langle$Fe$\rangle$ versus $\sigma_*$ and environment. The main findings are:

\begin{itemize}

\item All galaxies show a good correlation between Mgb/$\langle$Fe$\rangle$ and $\sigma_*$ ($r_S=0.76$), but have larger scatters in low $\sigma_*$ ($<80$km/s) bins.

\item For `old' (H$\beta<3$) low-$\sigma_*$ galaxies, their Mgb/$\langle$Fe$\rangle$ shows an obvious enhancement in high-local density regions and inner regions within galaxy groups. This trend is not clear for `young' (H$\beta>3$) and/or high-$\sigma_*$ ($>158$km/s)  galaxies. 

\item We do not find a systematic trend for the LSS environment or the central/satellite environment.  `Old' satellite galaxies, however, have the highest residual ($\delta\log$(Mgb/$\langle$Fe$\rangle$)) values in most $\sigma_*$ bins and galaxy regions.

\item `Old' galaxies with high formation redshift ($z_f>1.5$) have the highest residuals in all $\sigma_*$ bins and all galaxy regions, however, there is no clear systematic trend with formation redshift.

\item The Mgb/$\langle$Fe$\rangle$ gradient is close to 0 (flat) and depends on $\sigma_*$ but not obviously on environments. 
\end{itemize}

In conclusion, both the Mgb/$\langle$Fe$\rangle$ overall values and gradients mainly depend on galaxy properties such as stellar velocity dispersion. Environment plays a slight but obvious role in shaping the overall values of Mgb/$\langle$Fe$\rangle$; it has, however, no clear effect on Mgb/$\langle$Fe$\rangle$ gradients.

\section*{Acknowledgements}
This work is supported by the National Natural Science Foundation of China No. 11703036 (ZZ). 
This work is also partly supported by the National Key Basic Research and Development Program of China (No. 2018YFA0404501 to SM), by the National Science Foundation of China (Grant No. 11333003, 11390372 and 11761131004 to SM, and 11874057 to CLiu).
HW is supported by NSFC 11733004, 11522324 and 11421303.
The Science, Technology and Facilities Council is acknowledged for support through the Consolidated Grant Cosmology and Astrophysics at Portsmouth, ST/N000668/1.

This work makes use of data from SDSS-IV.
Funding for SDSS-IV has been provided by the Alfred P. Sloan Foundation and Participating Institutions. 
Additional funding towards SDSS-IV has been provided by the U.S. Department of Energy Office of Science. SDSS-IV
acknowledges support and resources from the Centre for High-Performance Computing at the University of Utah. The SDSS web site is www.sdss.org.

SDSS-IV is managed by the Astrophysical Research Consortium for the Participating Institutions of the SDSS Collaboration including the Brazilian Participation Group, the Carnegie Institution for Science, Carnegie Mellon University, the Chilean Participation Group, the French Participation Group, Harvard-Smithsonian Center for Astrophysics, Instituto de Astrof\'isica de Canarias, The Johns Hopkins University, Kavli Institute for the Physics and Mathematics of the Universe (IPMU) / University of Tokyo, Lawrence Berkeley National Laboratory, Leibniz Institut f$\ddot{\rm u}$r Astrophysik Potsdam (AIP), Max-Planck-Institut f$\ddot{\rm u}$r Astronomie (MPIA Heidelberg), Max-Planck-Institut f$\ddot{\rm u}$r Astrophysik (MPA Garching), Max-Planck-Institut f$\ddot{\rm u}$r Extraterrestrische Physik (MPE), National Astronomical Observatory of China, New Mexico State University, New York University, University of Notre Dame, Observat$\acute{\rm a}$rio Nacional / MCTI, The Ohio State University, Pennsylvania State University, Shanghai Astronomical Observatory, United Kingdom Participation Group, Universidad Nacional Aut$\acute{\rm o}$noma de M$\acute{\rm e}$xico, University of Arizona, University of Colorado Boulder, University of Oxford, University of Portsmouth, University of Utah, University of Virginia, University of Washington, University of Wisconsin, Vanderbilt University, and Yale University.

\appendix

\section{[$\alpha$/Fe]-$\sigma_*$ relation}
\label{appendix}
We fit the four measured spectral indices (H$\beta$, Mgb, Fe5270 and Fe5335) with the \citet{Thomas et al.2011} model. The \citet{Thomas et al.2011} model provides spectral indices in MILES resolution ($\sim 2.7$ \AA \, FWHM), which is similar to the SDSS MaNGA resolution. The model has a fine age grid: t = 0.1,0.2,0.4,0.6,0.8,1,2,3,4,5,6,7,8,9,10,11,12,13,14,15 Gyr; a relatively coarse metallicity grid: [Z/H] = -2.25, -1.35, -0.33, 0.0, 0.35, 0.67; and a very coarse $\alpha$ abundance grid: [$\alpha$/Fe] = -0.3, 0.0, 0.3, 0.5. 
Following \citet{Liu et al.2016}, we interpolate the model using a finer grid, i.e. using a step of 0.01 dex in the [Z/H] and [$\alpha$/Fe] grid. 

\begin{figure} 
\begin{center}
\includegraphics[scale=0.5]{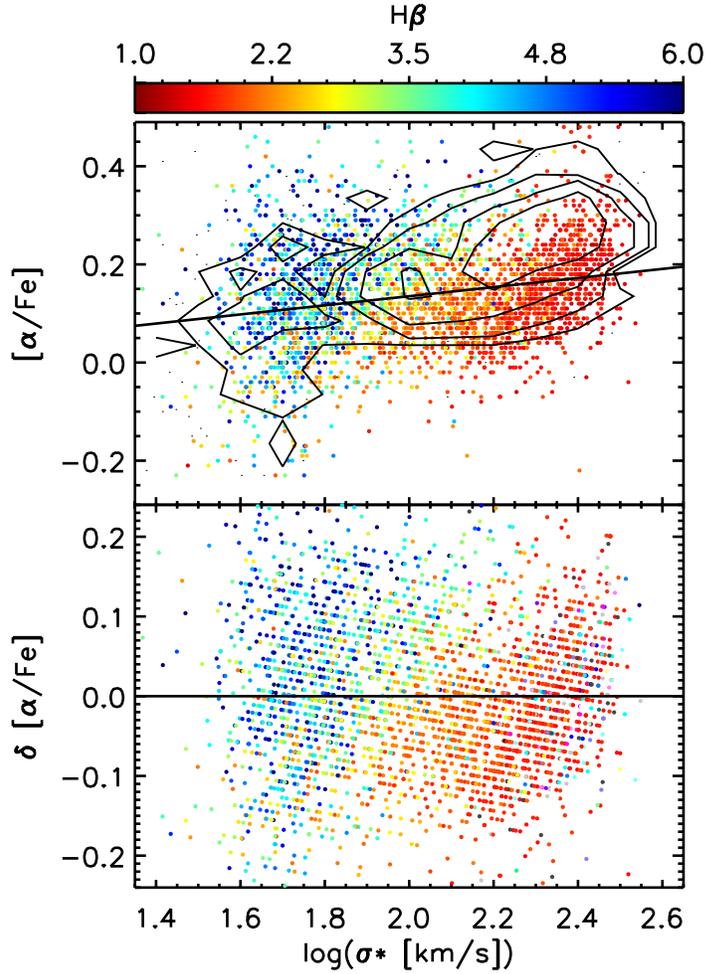}  
\caption{Upper panel: $\rm [\alpha/Fe]$ versus velocity dispersion for the central regions (within $\rm R_e/3$); the solid straight line is a linear fit to all the data points. Lower panel: the residuals, i.e. deviations of individual data points of the upper panel from the fitted straight line. Symbols are color-coded by H$\beta$. The black contours in the upper panel show measurements from the literature compiled by \citet{Liu et al.2016}. }
\label{a2fe_sigma_bulge_den}
\end{center}
\end{figure}

The fitted [$\alpha$/Fe] values for the central regions of our sample galaxies are presented in Fig. \ref{a2fe_sigma_bulge_den}. The [$\alpha$/Fe] also has a good correlation with $\sigma_*$, which is consistent with previous studies (cf. black contour in Fig. \ref{a2fe_sigma_bulge_den}), however, the [$\alpha$/Fe] - $\sigma_*$ relation has a smaller slope and a much larger scatter comparing to the Mgb/$\langle$Fe$\rangle$ - $\sigma_*$ relation.

\begin{figure} 
\begin{center}
\includegraphics[scale=0.6]{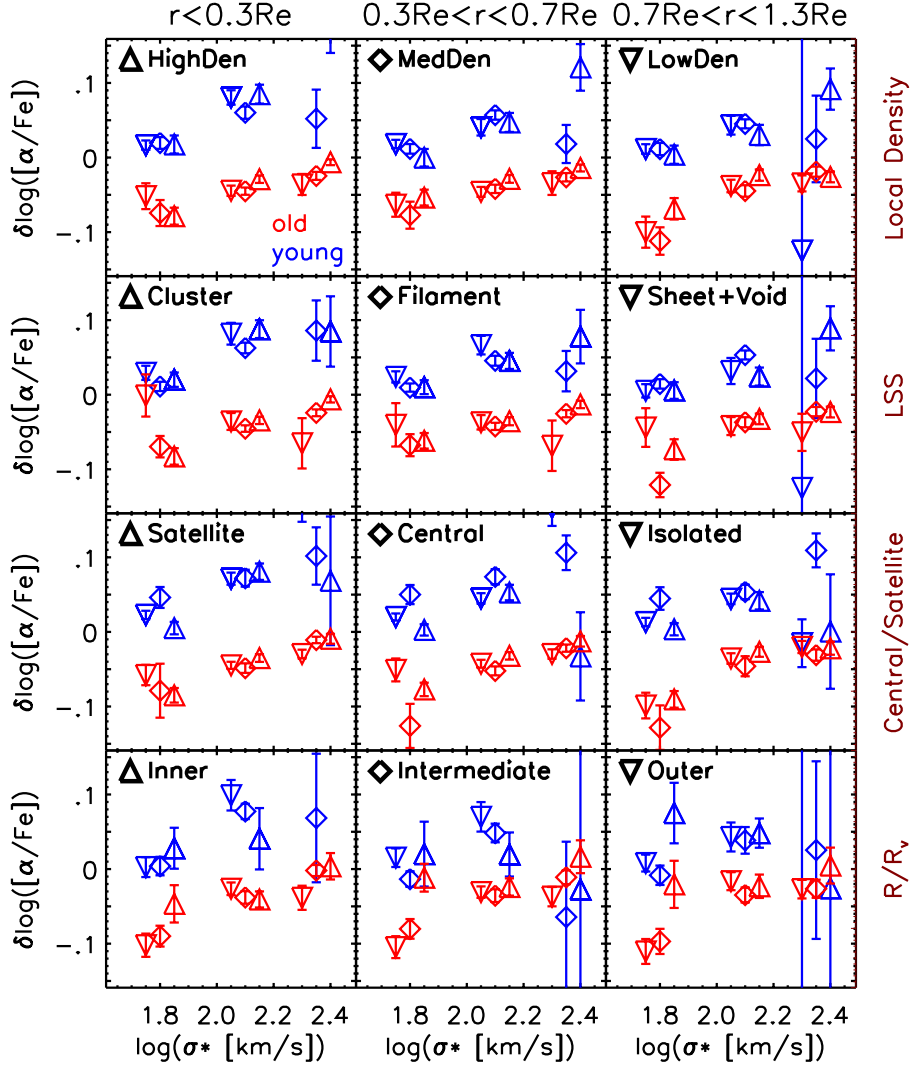}  
\caption{
\footnotesize
Deviation of $\rm [\alpha/Fe]$ from the  $\rm [\alpha/Fe]$-$\sigma_*$ relation for different parts of galaxies: central ($r<0.3R_e$, left column), intermediate ($0.3R_e<r<0.7R_e$, middle column), and outer ($0.7R_e<r<1.3R_e$, right column) and for different galaxy environment indicators: local density (top row), large scale structure (second row), isolated/central/satellite (third row) and radii to group centers (bottom row, satellite galaxies only). The plots are similar to Fig. \ref{sindx_env} but for $\rm [\alpha/Fe]$. The upward triangles, downward triangles and diamonds for each row are denoted in the legends of the plot. Red symbols are for galaxies with H$\beta<3$ (old) and blue symbols are for galaxies with H$\beta>3$ (young).}
\label{a2fe_env}
\end{center}
\end{figure}

We also show the environmental dependence of [$\alpha$/Fe] - $\sigma_*$ in Fig. \ref{a2fe_env}. It has much less environmental dependence comparing to the Mgb/$\langle$Fe$\rangle$ - $\sigma_*$ relation (Fig. \ref{sindx_env}). This could be due to the large scatter in [$\alpha$/Fe] derivation. 

\end{document}